\documentclass[lettersize, journal]{IEEEtran} 
\IEEEoverridecommandlockouts

\usepackage{cite}
\usepackage{amsmath, amssymb, amsfonts, mathtools}
\usepackage{algorithmic}
\usepackage{graphicx}
\usepackage{textcomp}
\usepackage{booktabs}
\usepackage{xcolor} \usepackage{relsize}
\def\BibTeX{{\rm B\kern-.05em{\sc i\kern-.025em b}\kern-.08em
    T\kern-.1667em\lower.7ex\hbox{E}\kern-.125emX}}

\usepackage{stfloats}

\graphicspath{{./Figures/}}
\newtheorem{Remark}{Remark}

\newtheorem{Lemma}{Lemma}
\newtheorem{Theorem}{Theorem}
\newtheorem{Corollary}{Corollary}

\newtheorem{proposition}{Proposition}

\DeclareMathAlphabet{\mathit}{OT1}{bch}{m}{it}

\DeclareFontFamily{U}{mathx}{}
\DeclareFontShape{U}{mathx}{m}{n}{<-> mathx10}{}
\DeclareSymbolFont{mathx}{U}{mathx}{m}{n}
\DeclareMathAccent{\widehat}{0}{mathx}{"70}
\DeclareMathAccent{\widecheck}{0}{mathx}{"71}

\newcommand{\src}{{\textnormal{S}}}
\newcommand{\des}{{\textnormal{D}}}
\newcommand{\ris}{{\textnormal{R}}}
\newcommand{\uav}{{\textnormal{U}}}

\newcommand{\AB}{\textnormal{AB}}

\newcommand{\nodeA}{\textnormal{A}}
\newcommand{\nodeB}{\textnormal{B}}
\newcommand{\PL}{\ell}

\newcommand{\snr}{{\gamma}}
\newcommand{\avgsnr}{\bar{\gamma}}

\newcommand{\atog}{{\mathsf{a2g}}}
\newcommand{\gtoa}{{\mathsf{g2a}}}
\newcommand{\etoe}{\mathtt{e2e}}
\newcommand{\var}{\textnormal{Var}}

\newcommand{\area}{{\mathcal{R}}}
\newcommand{\rwp}{{\mathtt{rwp}}}

\newcommand{\diag}{\textnormal{diag}}
\newcommand{\mean}{\textnormal{E}}

\newcommand{\real}{\textnormal{Re}}
\newcommand{\imag}{\textnormal{Im}}
\newcommand{\cov}{\textnormal{Cov}}

\newcommand{\refp}{{\mathtt{ref}}}

\newcommand{\deswp}{{\mathtt{des}}}
\newcommand{\srcwp}{{\mathtt{src}}}

\newcommand{\tsum}{\mathop{\textstyle\sum}}

\usepackage{geometry} 				
\begin{document}
\allowdisplaybreaks

\title{Ground-to-UAV and RIS-assisted UAV-to-Ground Communication Under Channel Aging: Statistical Characterization and Outage Performance}

\author{ 
	Thanh Luan~Nguyen, Georges~Kaddoum,~Tri~Nhu~Do,~Zygmunt~J.~Haas,

\vspace{-0.7cm}

    
	\thanks{T.-L. Nguyen and G.~Kaddoum are with the Department of Electrical Engineering, the \'{E}cole de Technologie Sup\'{e}rieure (\'{E}TS), Universit\'{e} du Qu\'{e}bec, Montr\'{e}al, QC H3C 1K3, Canada. 
    {G.~Kaddoum is also with Artificial Intelligence \& Cyber Systems Research Center, Lebanese American University} (emails: thanh-luan.nguyen.1@ens.etsmtl.ca, georges.kaddoum@etsmtl.ca).}

	\thanks{T. N. Do is with Electrical Engineering Department, Polytechnique Montr\'{e}al, Montr\'{e}al, QC, Canada (email: tri-nhu.do@polymtl.ca).}
	
	\thanks{Z. J. Haas is with Department of Computer Science, University of Texas at Dallas, TX 75080, USA, and also with the School of Electrical and Computer Engineering, Cornell University, Ithaca, NY 14853, USA (e-mail: haas@cornell.edu).}
}


\maketitle

\begin{abstract}
    This paper studies the statistical characterization of ground-to-air (G2A) and reconfigurable intelligent surface (RIS)-assisted air-to-ground (A2G) communications in RIS-assisted UAV networks under the impact of channel aging. 
    A comprehensive channel model is presented, which incorporates the time-varying fading, three-dimensional (3D) mobility, Doppler shifts, and the effects of channel aging on array antenna structures. 
We provide analytical expressions for the G2A signal-to-noise ratio (SNR) probability density function (PDF) and cumulative distribution function (CDF), demonstrating that the G2A SNR follows a mixture of noncentral $\chi^2$ distributions. The A2G communication is characterized under RIS arbitrary phase-shift configurations, showing that the A2G SNR can be represented as the product of two correlated noncentral $\chi^2$ random variables (RVs). Additionally, we present the PDF and the CDF of the product of two independently distributed noncentral $\chi^2$ RVs, which accurately characterize the A2G SNR's distribution. 
    Our paper confirms the effectiveness of RISs in mitigating channel aging effects within the coherence time. 
Finally, we propose an adaptive spectral efficiency method that ensures consistent system performance and satisfactory outage levels when the UAV and the ground user equipments are in motion.
\end{abstract}


\section{Introduction}
\IEEEPARstart{I}{ntegrated} terrestrial and non-terrestrial networks (TNTNs) have emerged as a pivotal innovation in the realm of wireless systems, particularly for the advent of six-generation (6G) networks \cite{GeraciCM2023, AzariCST2022, manzoor2024combined}.
    The traditional non-terrestrial networks (NTNs), previously limited to ground infrastructure coverage, have seen a myriad of new applications emerge in beyond fifth-generation (B5G) and 6G networks when integrated with terrestrial communication networks \cite{3gpp2018study}.
This integration is envisioned to offer widespread, seamless, and high-speed connectivity for a multitude of device types and applications, such as high-altitude platforms, unmanned aerial vehicles (UAVs), different types of ground base stations (BSs) and both stationary and mobile ground user equipments (GUEs).
    In this context, three-dimensional (3D) flying UAV platforms, such as Dà-Jiāng Innovations drones, play a critical role in enhancing the capabilities of TNTNs by providing flexible, mobile platforms that can be rapidly deployed to extend network coverage and capacity in disadvantaged or remote areas \cite{MozaffariCM2021, GeraciCST2022, manzoor2024combined}. 
    
Authorized by the Federal Aviation Administration (FAA), civilian drones are governed by a flight ceiling of $400$~feet ($\approx 120$~m) above ground level. 
    Additionally, these drones are permitted to fly at speeds of up to $100$~mph ($\approx 40$~m/s) \cite{copley2014faa}. To optimize their performance, UAVs can communicate over several frequency bands, including the $2$ GHz \cite{3GPP2019}, $5.8$ GHz, and $5030$-$5091$ MHz bands. 
    In the context of TNTNs, UAVs are recognized for their cost-effectiveness and adaptability to offer ubiquitous coverage to support a diverse range of applications \cite{FengIN2020}. 
Specifically, the integration of UAVs with cutting-edge technologies, such as Terahertz (THz) communication, massive multiple-input multiple-output (mMIMO) communications, and reconfigurable intelligent surfaces (RISs) are instrumental in facilitating high-performance wireless communications, even in disaster-stricken or remote areas \cite{YangTVT2020, YangTVT2022}.
    However, the precision of UAV tracking, and the ground-to-air (G2A) and air-to-ground (A2G) communications can be affected by position deviation issues due to wobbling and antenna jittering caused by body vibrations \cite{WangCL2024, YangTITS2023}.
Maintaining direct Line-of-Sight (LoS) to the GUE facilitates consistent connections, whereas Non-Line-of-Sight (NLoS) communication may require relay stations for reliable connectivity.
    Unfortunately, UAVs might operate in complex and unpredictable propagation environments, with A2G links occasionally blocked by trees or high-rise buildings \cite{GeraciCST2022}.
In addition, the mobility of UAVs introduces a phenomenon termed {\it channel aging} due to UAV’s variations in the G2A and A2G channels over time. 
    This phenomenon causes a mismatch between the actual channel state information (CSI) and the estimated CSI, rendering the estimated CSI outdated even if the channel acquisition was perfect, which degrades the performance of UAV-integrated wireless systems \cite{ChopraTWC2018, MozaffariCST2019, PapazafeiropoulosTVT2023, LiCL2023}.

Recent research works have widely studied the deployment of RISs in various environments, such as on the exteriors of high-rise buildings, to improve A2G communications \cite{BansalTC2023, FanTVT2023, NguyenTWC2024, AbualhayjaCL2024}.
    Using planar arrays of many low-cost passive elements, which can range from a few to thousands of elements \cite{DiamantiSMARTCOMP2021, DoTCOM2021, BjornsonCM2020}, RISs can intelligently manipulate the direction of the impinging electromagnetic waves, thus improving coverage and reliability, especially when direct LoS links are scarce.
However, RIS operation requires perfect CSI for optimal phase shift configuration, posing a challenge in dynamic communication environments like UAV-integrated systems. 
    Next, we shed light on some recent works that are relevant to our research.
    
    The study in \cite{LiCL2023} considered networks of cellular-connected UAV swarms and a massive MIMO BS, where the authors developed a closed-form expression for the signal-to-interference-plus-noise ratio that determines the uplink spectral efficiency (SE) of RIS-assisted A2G channel. 
In \cite{AbualhayjaCL2024}, the authors studied multi-hop multi-RIS-aided systems and proposed characterizing the true distribution of the RIS-assisted channel's magnitude by a double Nakagami-$m$ distribution. 
    This approach is similar to the generalized-$K$ distribution used to describe the behavior of a more straightforward system with a single RIS and a single-hop system in \cite{YangTVT2022}. 
Under the Rayleigh fading assumption, the authors in \cite{AlvaradoWCL2024} derived the SNR's exact distribution in RIS-assisted communication systems without LoS, following which the Central Limit Theorem (CLT) was adopted to approximate the derived results to noncentral $\chi^2$ distribution.
    In \cite{BansalTC2023}, which considered a single-antenna UAV serving multiple GUE with the aid of multiple RISs under independent and not necessarily identically distributed (non-IID) Rician fading with imperfect and outdated CSI, the authors applied the CLT to characterize the A2G signal-to-noise ratio (SNR) by a noncentral chi-squared ($\chi^2$) distribution.
While this noncentral $\chi^2$ distribution is mentioned in \cite{BansalTC2023} and \cite{AlvaradoWCL2024}, the degrees of freedom (d.o.f.) are set to $2$, which oversimplifies the distribution. Specifically, the noncentral $\chi^2$  distribution, with $2k$ d.o.f., noncentrality $2\lambda$, and scale $\frac{\avgsnr}{2}$, is equivalent to the generalized $\kappa$-$\mu$ distribution through a parameter transformation. This highlights a research gap that requires further study for more accurate characterization.
    In another approach, \cite{PapazafeiropoulosTVT2023} examined the effects of channel aging on RIS-enhanced massive MIMO systems with mobile UEs, where SE optimization under imperfect CSI was presented. The research also suggested that performance degradation due to channel aging can be mitigated by adjusting the frame duration or increasing the number of RIS elements.
Moreover, the authors in \cite{LuTC2024} evaluated the performance of RIS-assisted communications, accounting for hardware imperfections and channel aging due to GUE movement. 
    In \cite{LiWCL2024}, the authors considered a UAV-mounted RIS system and characterized the end-to-end (e2e) SNR using the $\alpha$-$\mu$ distribution, where the CLT is adopted for asymptotically large numbers of reflecting elements, e.g., up to $64^2$ elements. 
However, the works in \cite{PapazafeiropoulosTVT2023} and \cite{LuTC2024} did not consider UAV systems. In addition, although UAV systems are considered in \cite{AbualhayjaCL2024, BansalTC2023, LiWCL2024}, the impact of channel aging was not addressed in any of the references.
While RIS-assisted UAV wireless communication has gained significant attention in the literature, there is limited analysis of the statistical characterization of G2A and RIS-assisted A2G communications under channel aging. Moreover, most works do not take into account the size of the RISs when the number of reflecting element grows.
    This research gap has resulted in a lack of tools to facilitate the systematic design, such as how large can the SE be achieved without compromising the end-to-end outage probability (eOP) when the UAV and GUE are in motion.
    
This paper aims to fill the aforementioned gaps by studying the statistical characterization of G2A and RIS-assisted A2G communications under the impact of channel aging. 
    Specifically, we derive the exact distribution, including the probability density function (PDF) and cumulative distribution function (CDF), of the G2A SNR using Laplace transform and inverse Laplace transform. 
For the RIS-assisted A2G communication, we first derive the characteristic function (CF) of the A2G communication under arbitrary phase-shift configurations (PSCs). 
    We then propose characterizing the A2G channel as a complex Gaussian distribution conditioned on the estimated (i.e., delayed) CSI and RIS phase shifts.
The key contributions of this paper can be summarized as follows\footnote{Part of this paper was presented at the 2024 IEEE International Conference on Communications \cite{nguyen2024statistical}.}:
\vspace{-5pt}
\begin{itemize}
    \item We consider an RIS-assisted UAV system designed to accommodate the three-dimensional (3D) movements of the UAV and GUE. 
    Specifically, our analysis can be used for the BS and RIS with antenna array architectures, with uniform planar array (UPA) as a special case. As a result, our analysis is also applicable for large RISs, where the size of the RIS cannot be neglected. Moreover, we introduce a comprehensive channel model that accounts for time-varying fading, continuous-time mobility, Doppler frequency shifts, and channel aging.
    \item For the G2A communication, we derive a analytically tractable expression for the G2A SNR PDF while the delayed CSI-based maximal ratio combining (MRT) is adopted. Our findings reveal that the G2A SNR follows a mixture of noncentral $\chi^2$ distributions, which is later shown to be directly related to the $\kappa$-$\mu$-squared distribution. This important finding enables us to determine the exact G2A SNR CDF without further complex derivations.
    \item The unique technical contribution of our paper lies in our unified characterization of the A2G communication under arbitrary PSC. 
        Specifically, we derive the exact CF of the complex RIS-assisted A2G channel, conditioned on delayed CSI and RIS phase shifts. Then, the conditional A2G channel is found to follow a complex Gaussian distribution when the RIS has sufficiently large number of elements, denoted as $N$. 
    This leads us to a novel finding that the A2G SNR is characterized by the product of two correlated non-central $\chi^2$ RVs. 
        Simulation of Kullback–Leibler divergence between the exact and characterized A2G SNR is also presented to illustrate the accuracy of the proposed characterization.
    \item We tailor the method of moments to characterize the A2G SNR as the product of two independent noncentral $\chi^2$~RVs. Our approach includes a framework for matching central moments of the noncentral $\chi^2$ distribution, where the law of total cumulance (LTC) is utilized for this purpose. 
        Additionally, we introduce novel and exact PDF and CDF expressions for the product of two non-IID noncentral $\chi^2$ RVs to derive the A2G SNR PDF and CDF.
    Interestingly, our results show that the derived PDF/CDF is surprisingly accurate even with only $N = 2^2$ reflecting elements.
    \item For the system performance analysis, we examine the eOP under a fixed target SE and its asymptotic behavior. As a case study, we employ random mobility models for the GUE and UAV, specifically the random waypoint mobility (RWM) model and reference point group mobility (RPGM) model, to examine temporal fluctuations in the eOP. 
    Subsequently, we introduce an adaptive SE approach to maintain the eOP at satisfactory levels, ensuring a consistent and reliable connection as the GUE and UAV are in motion.
\end{itemize}

\noindent \textit{Notations}: $\mean[\cdot]$, $\var[\cdot]$, $\mu_3[\cdot]$ denotes expectation, variance, and third central moment operators, respectively; $X \mathop{=}^d Y$, $X \mathop{\approx}^d Y$, and $X \mathop{\to}^d Y$ indicate that the RVs $X$ and $Y$ are equal, approximately equal in distribution, and $X$ is matched to $Y$ in distribution, respectively.

    \vspace{-5pt}
\section{System Model}

\begin{figure}[!t]
    \centering
    \includegraphics[width = 0.7 \linewidth]{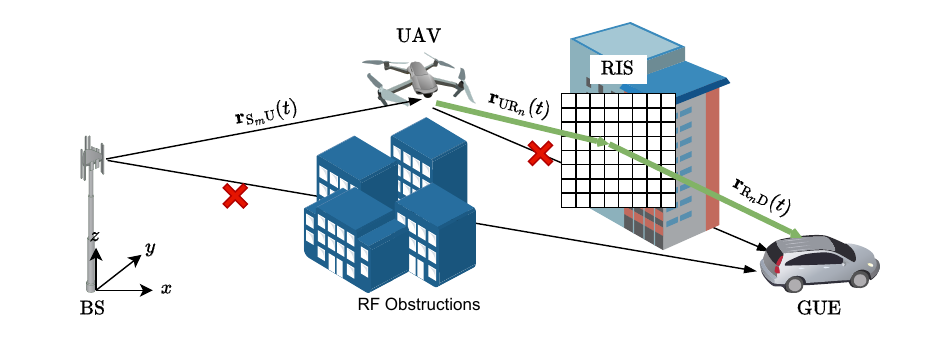}
    \caption{Illustrations of a G2A and RIS-assisted A2G wireless system, where the RIS is installed on houses and/or buildings.}
    \label{fig:systemModel}
    \vspace{-10pt}
\end{figure}

We consider a UAV-assisted terrestrial network, shown in Fig.~\ref{fig:systemModel}, where 
a single-antenna UAV is serving a single-antenna GUE by relaying the information from the $M$-antenna BS.
    In this system, the direct BS-GUE and UAV-GUE links are absent due to various Radio Frequency (RF) obstructions \cite{GeraciCST2022}. 
Hence, we consider a RIS with $N$ reflecting elements, installed on a building facade, to facilitate a virtual LoS link for the A2G communication \cite{BansalTC2023, FanTVT2023, NguyenTWC2024, AbualhayjaCL2024}.
    Hereafter, we use $\src$, $\src_m$, $\uav$, $\ris$, $\ris_n$, and $\des$ as subscripts to depict the BS, the BS's antenna $m \in {\cal M} \triangleq \{ 1, 2, \dots, M \}$, the UAV, the RIS, the reflecting (or RIS) element $n \in {\cal N} \triangleq \{ 1, 2, \dots, N \}$, and the GUE, respectively.
In addition, the transmission from BS to GUE is equally divided into two orthogonal time slots as follows. 
    During the first time slot, the BS transmits the unit-energy information signal $x_\src$ to the UAV. In the second time slot, the UAV decodes $x_\src$ and forwards the re-encoded version ${\widehat{x}_{\uav}}$ to the GUE with the help of the RIS. 
Moreover, we denote $\mathbf{p}_{\AB}(t) \triangleq \left[ x_{\AB}(t), y_{\AB}(t), z_{\AB}(t) \right]^{\sf T}$ as the 3D Cartesian coordinates of node $\nodeA$ with respect to node $\nodeB$, where $\nodeA, \nodeB \in \{ \src, \src_m, \uav, \ris, \ris_n, \des \}$.
    The conversion from the spherical coordinates $\left[ d_\AB(t), \theta_\AB(t), \varphi_\AB(t) \right]^{\sf T}$ to the 3D Cartesian coordinates is \cite{haslwanter20183d}
\begin{align}
\begin{bmatrix}
    x_\AB(t) \\
    y_\AB(t) \\
    z_\AB(t)
\end{bmatrix}
\!=\! d_\AB(t) 
\!
\begin{bmatrix}
    \cos\theta_\AB(t) \cos\varphi_\AB(t) \\
    \cos\theta_\AB(t) \sin\varphi_\AB(t) \\
    \sin\theta_\AB(t)
\end{bmatrix}\! \!\triangleq\! d_\AB(t) \mathbf{r}_\AB(t),
\label{eq:sph2cart}
\end{align}
where the unit vector $\mathbf{r}_\AB(t)$, i.e., $\Vert \mathbf{r}_\AB(t) \Vert^2 = 1$, represents the direction from node $\nodeA$ to node $\nodeB$, while $d_\AB(t)$, $\theta_\AB(t) \in \left[0, \pi\right]$, and $\varphi_\AB(t) \in [0, 2\pi]$ denote the distance~[m], the elevation angle [rad], and the azimuth angle [rad], between nodes $\nodeA$ and $\nodeB$, respectively. 
    The conversion from the Cartesian coordinates to the spherical coordinates is given by
\begin{align}
&\left[
    d_\AB, \theta_\AB, \varphi_\AB
\right]^{\sf T}
= \bigg[
    \sqrt{x^2_\AB + y^2_\AB + z^2_\AB},
\nonumber\\
&\qquad\quad
    \sin^{-1} \bigg( \frac{z_\AB}{\sqrt{x_\AB^2 + y_\AB^2 + z_\AB^2}} \bigg),
    \cos^{-1} \left( \frac{y_\AB}{x_\AB} \right)
\bigg]^{\sf T}. \label{eq:cart2sph}
\end{align}

We also denote $\mathbf{p}_{\nodeA}(t) \triangleq\! \left[ x_{\nodeA}(t), y_{\nodeA}(t), z_{\nodeA}(t) \right]^{\sf T}$ as node $\nodeA$'s Cartesian coordinates at time $t$ with respect to the origin in a predefined global coordinates system, thus ${\mathbf{p}_{\AB}(t) \!=\! \mathbf{p}_{\nodeB}(t) - \mathbf{p}_{\nodeA}(t)}$. 

\vspace{-10pt}
\subsection{Time-Varying Fading Channel}

    We denote the time-varying channel coefficient from node $\nodeA$ to node $\nodeB$ at time $t$ as $c_{\AB}(t) \triangleq \sqrt{\PL_\AB(t)} h_{\AB}(t)$, where $\PL_\AB(t)$ is the path loss component and $h_{\AB}(t)$ is the small-scale fading component.
Moreover, we consider that all channels experience Rician fading comprising a line-of-sight (LoS) path and multiple non-LoS (NLoS) paths. Specifically, the channel vectors are modeled as $\mathbf{h}_{\AB}(t)
    =   \frac{\sqrt{K_{\AB}(t)}}{\sqrt{K_{\AB}(t) + 1}} 
        \bar{\mathbf{h}}_{\AB}(t)
    + \frac{\widetilde{\mathbf{h}}_{\AB}(t)}{\sqrt{K_{\AB}(t) + 1}}$ \cite{WuTVT2020}, where $\bar{\mathbf{h}}_{\AB}(t)$ represents the LoS component and $\widetilde{\mathbf{h}}_{\AB}(t)$ represents the NLoS components whose entries are independent and identically distributed (IID) complex Gaussian RVs with zero mean and unit variance. 
    The Rician factor is modeled as a function of the elevation angle as $K_\AB(t) = K_0 e^{\frac{2}{\pi} \theta_\AB(t) \ln\frac{K_\pi}{K_0}}$ for LoS scenarios, where $K_0$ and $K_\pi$ are environmental coefficients. It is noted that $K_\AB(t) = 0$ indicates the NLoS scenarios, where the channel experience Rayleigh fading.
When antenna array architectures are employed at the BS, we rewrite $\bar{\mathbf{h}}_{\AB}(t)$ as \cite[Eq. (5)]{XuTVT2023}
\begin{align}
\left[ \bar{\mathbf{h}}_{\src\uav}(t) \right]_m = 
    \exp\Big[ j 2 \pi t \Delta \textrm{f}_{\src_m\uav}(t) \Big]
    \sqrt{M} \boldsymbol {\alpha}_{\src}(\mathbf{r}_{\src\uav}(t)) ,
\end{align}
where $\Delta \textrm{f}_{\src_m\uav}(t)$ is the Doppler frequency shift in the LoS path induced by the movement of the UAV and $\boldsymbol{\alpha}_{\src}(\mathbf{r})$ is the normalized array response or normalized steering vector in the direction $\mathbf{r} \in \mathbb{R}^3$. 
    In particular, the vector $\mathbf{r} = \mathbf{r}_{\src\uav}(t)$ specifies the direction of the waveform departing from the BS toward the UAV, or the so-called direction of departure (DoD).
The array steering vector $\boldsymbol{\alpha}_{\src}(\mathbf{r})$ in the direction $\mathbf{r} \in \mathbb{R}^3$ can be expressed as \cite[Eq. (7.13)]{bjornson2017massive}
\begin{align}
\boldsymbol{\alpha}_{\src}(\mathbf{r})
    =
    \Big[ 
        e^{j \frac{2\pi}{\lambda} \mathbf{r}^{\sf T} \mathbf{p}_{\src_1 \src_0} },  \dots, e^{j \frac{2\pi}{\lambda} \mathbf{r}^{\sf T} \mathbf{p}_{\src_M \src_0} }
    \Big]^{\sf T} \Big/ \sqrt{M},
\label{prop:ary_steer_BS}
\end{align}
where $\mathbf{p}_{\src_m\src_0}$ denotes the position of the $m^{th}$ antenna with respect to the center of the antenna platform. Here, the wave vector in \cite[Eq. (7.13)]{bjornson2017massive} relates to the direction vector as $\mathbf{k} = \frac{2\pi}{\lambda} \mathbf{r}$.

When the BS is equipped with an UPA in the $yz$-plane with $M_{\textrm{H}}$,  $M_{\textrm{V}}$ and $d_{\textrm{H}}$, $d_{\textrm{V}}$ being the number of elements and element spacing on the $y$~(horizontal) axis, and the number of elements and element spacing on the $z$~(vertical) axis, respectively, we have 
$\mathbf{p}_{\src_m \src_0} = \left[ 0, (i_{\src_m}-1) d_{\textrm{H}}, (j_{\src_m}-1) d_{\textrm{V}} \right]^{\sf T}$, for $m \in {\cal M}$, where $i_{\src_m} = 1, \dots, M_{\textrm{H}}$ and $j_{\src_m} = 1, \dots, M_{\textrm{V}}$ are the corresponding horizontal and vertical subscripts of the $M_{\textrm{H}} \times M_{\textrm{V}}$ matrix, and where $M_{\textrm{H}} M_{\textrm{V}} = M$. Specifically, $i_{\src_m}$ and $j_{\src_m}$ are determined as $\left( i_{\src_m}, j_{\src_m} \right) = \left( \operatorname{mod}\left(m-1, M_{\textrm{H}} \right) + 1, \left\lfloor (m-1)/M_{\textrm{H}}\right\rfloor + 1 \right) \triangleq \textrm{ind2sub}\left( [ M_{\textrm{H}}, M_{\textrm{V}}], m \right)$ \cite{bjornson2017massive}. 

Following \eqref{prop:ary_steer_BS}, we rewrite the LoS component of the UAV-RIS and RIS-GUE channel vectors as
\begin{align}
\left[ \bar{\mathbf{h}}_{\uav\ris}(t) \right]_n
    &=  \exp\Big[ j 2 \pi t \Delta \textrm{f}_{\uav\ris_n}(t) \Big] 
    \sqrt{N} \boldsymbol {\alpha}_{\ris}(-\mathbf{r}_{\uav\ris_n}(t)), \\
\left[ \bar{\mathbf{h}}_{\ris\des}(t) \right]_n 
    &=  \exp\Big[ j 2 \pi t \Delta \textrm{f}_{\ris_n\des}(t) \Big] 
    \sqrt{N} \boldsymbol {\alpha}_{\ris}(\mathbf{r}_{\ris_n\des}(t)),
\end{align}
where $\Delta f_{\uav\ris_n}(t)$ and $\Delta f_{\ris_n\des}(t)$ are the Doppler frequency shifts in the LoS path induced by the movement of the UAV and the GUE, respectively, and $-\mathbf{r}_{\uav\ris_n}(t)$ is the direction of the waveform departing from the UAV and arriving at the $n^{th}$ element of the RIS. The direction $-\mathbf{r}_{\uav\ris_n}(t)$ is also referred to as the direction of arrival (DoA). Moreover, $\mathbf{r}_{\ris_n\des}(t)$ is the DoD of the waveform departing from the $n^{th}$ element of the RIS toward the GUE, and $\boldsymbol{\alpha}_{\ris}(\mathbf{r})
    =   
    \left[ 
        e^{j \frac{2\pi}{\lambda} \mathbf{r}^{\sf T} \mathbf{p}_{\ris_1 \ris_0} }, \dots, 
        e^{j \frac{2\pi}{\lambda} \mathbf{r}^{\sf T} \mathbf{p}_{\ris_N \ris_0} }
    \right]^{\sf T} \Big/ \sqrt{N}$.
When the RIS has an UPA architecture in the $yz$-plane with $N_{\textrm{H}}$, $N_{\textrm{V}}$, and $d_{\textrm{H}}$, $d_{\textrm{V}}$ being the number of elements and element spacing on the horizontal axis and the number of elements and element spacing and vertical axis, respectively, we have 
$\mathbf{p}_{\ris_n \ris_0} = \left[ 0, (i_{\ris_n}-1) d_{\textrm{H}}, (j_{\ris_n}-1) d_{\textrm{V}} \right]^{\sf T}$ for $n \in {\cal N}$, where, for $i_{\ris_n} = 1, \dots, N_{\textrm{H}}$ and $j_{\ris_n} = 1, \dots, N_{\textrm{V}}$ and $\left( i_{\ris_n}, j_{\ris_n} \right)^{\sf T} = \textrm{ind2sub}\left( [ N_{\textrm{H}}, N_{\textrm{V}}], n \right)$.

\vspace{-10pt}
\subsection{Continuous-Time Mobility Model}

The GUE's motion with respect to the origin can be modeled by the change of its positions over time as ${\mathbf{p}_{\des}(t, \tau) = \mathbf{p}_{\des}(t+\tau) - \mathbf{p}_{\des}(t)} \in \mathbb{R}^3$ \cite{haslwanter20183d}, where $\tau$ [sec] is the delayed duration from the initial observation instant $t$ [sec]. 
The GUE's motion can also be determined based on its average velocity over a period $\tau$, denoted as $\mathbf{v}_{\des}(t, \tau)$, as
\begin{align}
\mathbf{p}_{\des}(t, \tau) 
    =  \mathbf{p}_{\des}(t) + \tau \mathbf{v}_{\des}(t, \tau).
\end{align}

In various communication scenarios, the UAV is required to move following the GUE's movement to ensure uninterrupted connectivity.
    In this case, the UAV's motion is a function of the GUE's motion, and is modeled as
\begin{align}
\mathbf{p}_{\uav}(t, \tau)
    =   \mathbf{p}_{\des}(t, \tau) + \Delta\mathbf{p}_{\uav}(t, \tau),
\label{eq:uav_motion_rpgm}
\end{align}
where $\Delta\mathbf{p}_{\uav}(t, \tau) \in \mathbb{R}^3$ represents the UAV's motion with respect to the GUE's motion.

The mobility of the UAV and GUE causes the A2G channel to vary over time, which leads to the channel aging phenomenon \cite{ZhangTVT2023, ZhengTWC2021, ChopraTWC2018}. Channel aging causes the CSI to become outdated due to the delay between the estimation time and the CSI transmission time. Moreover, the velocity vectors of the UAV and the GUE, i.e., $\mathbf{v}_{\uav}(t)$ and $\mathbf{v}_{\des}(t)$, also vary over time, which results in a time-varying Doppler frequency.

\vspace{-10pt}
\subsection{Time-Varying Doppler Frequency}

If nodes $\nodeA$ and $\nodeB$ move with instantaneous velocities $\mathbf{v}_{\nodeA}(t) \in \mathbb{R}^3$ and $\mathbf{v}_{\nodeB}(t) \in \mathbb{R}^3$, respectively, the Doppler shift along the LoS path is expressed as
\begin{align}
\Delta \textrm{f}_{\AB}(t) = \frac{f_c}{c} \left( \mathbf{v}_{\nodeB}(t)-\mathbf{v}_{\nodeA}(t) \right)^{\sf T} \mathbf{r}_{\AB}(t), \label{eq:dopplerFreq_AB}
\end{align}
where $\mathbf{r}_{\AB}(t)$ is determined in \eqref{eq:sph2cart}. We note that $\mathbf{v}_{\nodeA}(t) = 0$ (or $\mathbf{v}_{\nodeB}(t) = 0$) implies that node $\nodeA$ (or $\nodeB$) is standing still.

We note that \eqref{eq:dopplerFreq_AB} is applicable to 3D mobility scenarios, with one-dimensional (1D) mobility being a special case. In this context, the term $\mathbf{v}_{\nodeB}(t) - \mathbf{v}_{\nodeA}(t) \triangleq \mathbf{v}_{\AB}(t)$ indicates the relative velocity of node $\nodeB$ with respect to node $\nodeA$. 
    Moreover, \eqref{eq:dopplerFreq_AB} can be equivalently represented by one-dimensional (1D) mobility as $\Delta \textrm{f}_{\AB}(t) = \pm \frac{f_c}{c} \left\Vert \mathbf{v}_{\nodeB}(t) - \mathbf{v}_{\nodeA}(t) \right\Vert \cos\theta_\AB(t)$, where $\theta_\AB(t)$ is determined in \eqref{eq:cart2sph} and the $\pm$ sign is determined by whether the node $\nodeA$ is moving towards ($+$) or away ($-$) from node $\nodeB$, respectively. 
Based on \eqref{eq:dopplerFreq_AB}, we formulate the maximum Doppler frequency shift across the NLoS components as
\begin{align}
\Delta \textrm{f}^{\textrm{max}}_{\AB}(t) 
    = \frac{f_c}{c} \left\Vert \mathbf{v}_{\nodeB}(t) - \mathbf{v}_{\nodeA}(t) \right\Vert.
\end{align}

\subsection{Channel Aging Model}
\label{sec:channel_aging}

    To measure the CSI aging caused by the Doppler effect, correlation metrics are used to determine the time-varying property of the CSI.
Specifically, the channel state at a discrete time instant $t_k$ can be modeled as \cite{ZhengTWC2021, ChopraTWC2018}
\begin{align}
\mathbf{h}_{\AB}[t_k]
    = \rho_\AB[t_k] \mathbf{h}_{\AB}[0] + \bar{\rho}_{\AB}[t_k] \mathbf{f}_{\AB}[t_k],
\label{eq:aging_model}
\end{align}
where ${\mathbf{f}_{\AB}[t_k] \mathop{=}^d \mathcal{CN}({\bf 0}, \mathbf{I})}$ represents the innovation component, $\rho_{\AB}[t_k] \in [-1, 1]$ is the temporal  correlation coefficient between the channel realizations at time instances $0$ and $k$, and $\bar{\rho}_{\AB}[t_k] = \sqrt{1-\rho_\AB^2[t_k]}$. 
As in \cite{ChopraTWC2018, ZhengTWC2021}, we consider that the channels evolve according to Jakes’ model, thus
\begin{align}
    \rho_{\AB}[t_k] = J_0\Big( 2\pi t_k T_s \Delta \textrm{f}^{\textrm{max}}_{\AB}(t_k T_s) \Big),
\end{align}
where $J_0(x)$ is the zeroth-order Bessel function of the first kind and $T_s$ [sec] is the channel sampling duration. 
Since $\rho_{\AB}[t_k]$ is a Bessel function, either change in the UAV velocity, GUE velocity, or a higher sample index $t_k$ will yield a non-monotonic decrease in the correlation coefficient which oscillates around zero with decreasing magnitude. 

Considering that the channels' estimates at time instant $\tau$ are perfect and that these estimates are used as the initial states to get estimates of the channels at all other time instants, then the  channels' estimates at later time instants will be inaccurate \cite{ZhengTWC2021, PapazafeiropoulosTVT2023}. 
    Thus, the current channel at time instant $t_k$ can be expressed in terms of the estimated channel at the $\tau^{th}$ time instant as \cite{ZhengTWC2021}
\begin{align}
\mathbf{h}_{\AB}[t_k] 
    =  \rho_{\AB}[\tau-t_k] {\mathbf{h}}_{\AB}[\tau] + \bar{\rho}_{\AB}[\tau-t_k] \mathbf{z}_{\AB}[t_k], 
\label{eq:hab_aging_estimate_a}
\end{align}
where $\mathbf{z}_{\AB}[t_k] \sim \mathcal{CN}(\mathbf{0}, \mathbf{I})$ denotes the independent innovation component that correlates $\mathbf{h}_{\AB}[t_k]$ and the (outdated) estimated CSI ${\mathbf{h}}_{\AB}[\tau]$.     
Hereafter, we drop the time indices and denote the delayed CSI as ${\widehat{\mathbf{h}}_{\AB} = {\mathbf{h}}_{\AB}[\tau]}$ for convenience.

\vspace{-10pt}   
\subsection{Ground-to-Air Signal-to-Noise Ratio}

In the first time slot, the BS applies the beamforming vector $\mathbf{w}_{\src\uav}$ to steer the desired signal $x_\src$ to the UAV, where $\textnormal{E}[|x_{\src}|^2] = 1$. Hence, the received signal at the UAV is obtained as
\begin{align} \label{Eq10}
y_{\uav} = 
    \sqrt{P_\src \PL_{\src\uav}} 
    \left[
        (\rho_{\src\uav} \widehat{\mathbf{h}}_{\src\uav} 
            + \bar{\rho}_{\src\uav} \mathbf{z}_{\src\uav})^{\sf H} \mathbf{w}_{\src\uav} 
    \right] x_{\src}
    + n_\uav,
\end{align}
where $P_\src$ [W] denotes the BS's transmit power and $n_{\uav}$ is the additive white Gaussian noise (AWGN) with zero mean and variance~$\sigma^2_{\uav}$ [W] at the UAV.
    We assume that $\PL_{\src_1\uav} \approx \PL_{\src_2\uav} \approx \dots \approx \PL_{\src_M\uav} = \PL_{\src\uav}$ since the maximum dimension of the BS's antenna array is significantly smaller than its distance to the UAV.
Moreover, we consider that the BS adopts maximal ratio transmission (MRT) using the estimated CSI, where ${\mathbf{w}_{\src\uav} = \frac{\widehat{\mathbf{h}}_{\src\uav}}{\left\Vert\widehat{\mathbf{h}}_{\src\uav}\right\Vert}}$. 
Hence, the G2A SNR under delayed CSI-based MRT is obtained~as
\begin{align}
\snr_{\gtoa}
    =  \frac{P_\src\PL_{\src\uav}}{\sigma_\uav^2}
    \left| 
        (\rho_{\src\uav} \widehat{\mathbf{h}}_{\src\uav} + \bar{\rho}_{\src\uav} {\mathbf{z}}_{\src\uav} )^{\sf H} \widehat{\mathbf{h}}_{\src\uav}
    \right|^2
    \left\Vert \widehat{\mathbf{h}}_{\src\uav} \right\Vert^{-2}. \label{Eq11}
\end{align}

\vspace{-10pt}
\subsection{Air-to-Ground Signal-to-Noise Ratio}

In the second time slot, the UAV decodes $x_\src$ and forwards the re-encoded version ${\widehat{x}_{\uav}}$ to the RIS.
    Next, the RIS reflects the signal $\widehat{x}_{\uav}$ from the UAV to the GUE by intelligently adjusting its phase-shift matrix.
Hence, the received signal at the GUE is obtained~as
\begin{align}
y_{\des}
    =  \sqrt{P_{\uav}} 
    \left( 
        \sum_{n=1}^{N} \sqrt{{\PL}_{\ris_n\des}} h_{\ris_n\des} \beta_n \sqrt{{\PL}_{\uav\ris_n}} h_{\uav\ris_n} e^{j\vartheta_{\ris_n}}
    \right)
    \widehat{x}_\uav + n_\des,
\end{align}
where $P_{\uav}$ [W] is the UAV's transmit power, $n_\des$ is the AWGN with zero mean and variance $\sigma_\des^2$ [W] at the GUE, $\beta_n \in \mathbb{R}$ and ${ \vartheta_{\ris_n} \!=\! [0,2\pi) }$ are the amplitude reflection coefficient and the phase-shift value of the reflecting element ${ n \in {\cal N} }$, respectively \cite{ZhangTVT2023, BjornsonCM2020}. 
    In practice, the number of phase-shifts is limited and constrained by the phase-shift resolution, denoted by ${Q \triangleq 2^b}$, where $b$ is the number of quantization bits and the value of a phase shift belongs to the set ${\cal Q} = \left\{0, \frac{2\pi}{Q}, \frac{4\pi}{Q},...,\frac{2\pi(Q-1)}{Q} \right\}$ \cite{Huang_TWC_2019}. 
Hence, the SNR received at the GUE is obtained as
\begin{align} 
\snr_{\atog}
    &=  \frac{P_\uav}{\sigma_\des^2}
    \left|
        \sum_{n=1}^{N} \sqrt{{\PL}_{\ris_n\des}} h_{\ris_n\des} \beta_n \sqrt{{\PL}_{\uav\ris_n}} h_{\uav\ris_n} e^{j\vartheta_{\ris_n}}
    \right|^2 \label{eq:snr_a2g_1} \\
    &\mathop{\approxeq}\limits^{(a)}
    \avgsnr_{\atog}
    \left|
        \mathbf{h}_{\ris\des}^{\sf H} \boldsymbol{\Theta}_{\ris} \mathbf{h}_{\uav\ris}
    \right|^2, \forall n \in {\cal N}, \label{eq:snr_a2g}
\end{align}
where $\boldsymbol{\Theta}_{\ris} \triangleq \diag\left( \beta_1 e^{j\vartheta_{\ris_1}}, \dots, \beta_N e^{j\vartheta_{\ris_N}} \right)$ represents the phase-shift matrix, 
    $\avgsnr_{\atog} \triangleq \frac{P_\uav}{\sigma_\des^2} \ell_{\uav\ris_n} \ell_{\ris_n\des}$ for any $n \in {\cal N}$ when the RIS's maximum dimension being considerably smaller than the distances $d_{\ris_n\des}$ and $d_{\uav\ris_n}$ \cite{Huang_TWC_2019, DoTCOM2021}. It is noted that our analysis is applicable when the RIS dimensions are non-negligible. 
    In this case, we define $\ell_\atog \triangleq \frac{1}{N} \left( \sum_{n=1}^{N} \sqrt{\ell_{\ris_n\des}} \beta_n \sqrt{\ell_{\uav\ris_n}} \right)^2$ as the average effective A2G path loss, and the terms $\beta_n$ and $\avgsnr_{\atog}$ must be modified to $\frac{\sqrt{\ell_{\ris_n\des}} \beta_n \sqrt{\ell_{\uav\ris_n}}}{\sqrt{\ell_\ris}}$ and $\frac{P_\uav \ell_\atog}{\sigma_\des^2}$, respectively.

\vspace{-5pt}
\section{Statistical Characterization of G2A and RIS-assisted A2G Communication}

In this section, we study the statistical characteristics of the G2A and A2G communications. Specifically, we derive the distribution, i.e., the PDF and CDF, of the G2A SNR. For the A2G communication, we propose to characterize the A2G channel distribution conditioned on the estimated CSI. Then, the PDF and CDF of the A2G SNR are derived.
    Moreover, we focus on the distributions of the LoS communication scenario since the NLoS communication scenario can be analogously obtained by setting the Rician-$K$ factors to zeros \cite{MozaffariCST2019}.

\vspace{-5pt}
\subsection{Statistical Characterization of G2A Communication}

We denote $h_{\src\uav} \triangleq (\rho_{\src\uav} \widehat{\mathbf{h}}_{\src\uav} + \bar{\rho}_{\src\uav} {\mathbf{z}}_{\src\uav} )^{\sf H} {\mathbf{w}}_{\src\uav}$ as the effective G2A small-scale fading channel.
    For practical purposes, we consider a reasonable assumption that ${\mathbf{w}}_{\src\uav}$ and ${\mathbf{z}}_{\src\uav}$ are statistically independent. 
Subsequently, $h_{\src\uav}$ is characterized as $h_{\src\uav} \mathop{=}^d \mathcal{CN}\left( \rho_{\src\uav} {\mathbf{w}}_{\src\uav}^{\sf H} \widehat{\mathbf{h}}_{\src\uav}, \bar{\rho}_{\src\uav}^2 \right)$. 
    We find that $h_{\src\uav}$ exhibits the properties of the Rician fading, where $\rho_{\src\uav} {\mathbf{w}}_{\src\uav}^{\sf H} \widehat{\mathbf{h}}_{\src\uav}$ and $\bar{\rho}_{\src\uav}^2$ represent the effective LoS component and the average power of the NLoS components, respectively.
Applying the delayed CSI-based MRT beamforming, $h_{\src\uav}$ is further characterized~as
\begin{align}
h_{\src\uav}
    \mathop{=}^d \mathcal{CN}\Big( 
        \rho_{\src\uav} \left\Vert \widehat{\mathbf{h}}_{\src\uav} \right\Vert, \bar{\rho}_{\src\uav}^2 
    \Big).
\label{eq:eff_g2a_channel}
\end{align}

Under the delayed CSI-based MRT beamforming, ${\mathbf{w}}_{\src\uav}^{\sf H} \widehat{\mathbf{h}}_{\src\uav}$ follows the noncentral chi-square ($\chi^2$) distribution with scale factor $\frac{1}{2(K_{\src\uav}+1)}$, $2M$ degrees of freedom (d.o.f.), and noncentrality parameter $2MK_{\src\uav}$.
    Such a fading falls into the category of fluctuating LoS fading \cite{AbdiTWC2003}. However, \cite{AbdiTWC2003} only considers that the magnitude of the LoS component follows a Nakagami-$m$ distribution, forming the Gamma-shadowed Rician fading channel. Hereafter, we use the notation $\widetilde{\chi}_{k}^2\left( \lambda \right)$ to specify the noncentral $\chi^2$ RV with scale $\frac{1}{2}$, $2 k$ d.o.f., and noncentrality parameter $2\lambda$.
The PDF and CDF of $\widetilde{\chi}_k^2\left( \lambda \right)$ are given by \cite{Mathai1992}
\begin{align}
f_{\widetilde{\displaystyle \chi}^2}(k, \lambda; x) 
    &=   e^{-x-\lambda} \left( \frac{x}{\lambda} \right)^{\frac{k-1}{2}} I_{k-1}\left( 2\sqrt{\lambda x} \right),~ x > 0, \\
F_{\widetilde{\displaystyle \chi}^2}(k, \lambda; x) 
    &=   1 - Q_k\left( \sqrt{2\lambda}, \sqrt{2 x} \right),~ x > 0, \label{eq:cdf_noncentralChi2}
\end{align}
where $Q_k(a, b) = a^{1-k} \int_b^\infty x^k e^{-\frac{x^2+a^2}{2}} I_{v-1}(ax) \mathrm{d} x$ denotes the Marcum Q-function. For the noncentral $\chi^2$ distribution with scale factor $\avgsnr$, denoted as $\avgsnr \widetilde{\chi}_{k}^2\left( \lambda \right)$, we adopt the scaling property of PDFs and CDFs as $f_{\avgsnr \widetilde{\displaystyle \chi}^2}(k, \lambda; x) = \frac{1}{\avgsnr} f_{ \widetilde{\displaystyle \chi}^2}(k, \lambda; \frac{x}{\avgsnr})$ and 
$F_{\avgsnr \widetilde{\displaystyle \chi}^2}(k, \lambda; x) = F_{ \widetilde{\displaystyle \chi}^2}(k, \lambda; \frac{x}{\avgsnr})$, respectively, for $\avgsnr > 0$.
\begin{Remark}
\label{rem:scale_NCCS}
The distribution of $\avgsnr \widetilde{\chi}_{k}^2\left( \lambda \right)$ is directly related to the $\kappa$-$\mu$-squared distribution with fading parameters $(\Omega, \kappa, \mu)$. Specifically, the two distributions can be transformed into each other by setting the parameters as follows:

    + To obtain the $\kappa$-$\mu$-squared distribution from the noncentral $\chi^2$ distribution, set $\avgsnr = \frac{\Omega}{\mu (1+\kappa)}$, $\lambda = \mu \kappa$, and $k = \mu$.
    
    + To obtain the noncentral $\chi^2$ distribution from the $\kappa$-$\mu$-squared distribution, set $\mu = k$, $\kappa = \frac{\lambda}{k}$, and $\Omega = \avgsnr(k + \lambda)$.

%
\end{Remark}

\begin{Theorem} \label{theo:pdf_cdf_g2asnr}
Under the delayed CSI-based MRT, the PDF of the G2A SNR is formulated as
\begin{align}
f_{\snr_{\gtoa}}(x)
    &=  e^{ -M \frac{K_{\src\uav} \rho_{\src\uav}^2}{K_{\src\uav} \bar{\rho}_{\src\uav}^2+1}
    -\frac{(K_{\src\uav}+1)x}{\avgsnr_{\gtoa} (K_{\src\uav} \bar{\rho}_{\src\uav}^2 + 1) } }
    \left(
        \frac{K_{\src\uav}+1}{K_{\src\uav} \bar{\rho}_{\src\uav}^2+1}
    \right)^M
    \nonumber\\
    &\quad\times
    \sum_{i=0}^{M-1}
    \binom{M-1}{i}
    \frac{(\bar{\rho}_{\src\uav}^2)^{i} ({\rho}_{\src\uav}^2)^{M-i-1}}{\avgsnr_{\gtoa}^{M-i}}
    \left( \frac{x}{\Xi_{\uav}} \right)^{\frac{M-i-1}{2}}
    \nonumber\\
    &\qquad\qquad\times
    I_{M-i-1} \left(
        \frac{ 2 \sqrt{\Xi_{\uav} x} }{K_{\src\uav} \bar{\rho}_{\src\uav}^2+1}
    \right),~ x > 0,
\label{eq:pdf_snrG2a}
\end{align}
where $\Xi_{\uav} \triangleq M K_{\src\uav} \rho_{\src\uav}^2 (K_{\src\uav} +1) \avgsnr_{\gtoa}^{-1}$ and $\bar{\gamma}_{\src\uav} = \frac{P_\src\PL_{\src\uav}}{\sigma_\uav^2}$. 
%
\end{Theorem}

\begin{IEEEproof}
For convenience, the proof of \eqref{eq:pdf_snrG2a} is presented in Appendix \ref{apx:proof_g2aSNR}, which involves many mathematical manipulations based on the Laplace transform of $\snr_{\gtoa}$.
\end{IEEEproof}

%

\begin{Corollary}
\label{cor:cdf_snrG2a}
Based on \eqref{eq:pdf_snrG2a}, we find that the G2A SNR under the delayed CSI-based MRT follows a mixture of $M$ noncentral $\chi^2$ distributions, where the $m^{th}$ noncentral $\chi^2$ distribution is identically distributed to $\avgsnr \widetilde{\chi}_{M-m}(\lambda)$, for $m = 0, 1, \dots, M - 1$, where 
    $\lambda \triangleq M \frac{K_{\src\uav} {\rho}_{\src\uav}^2}{K_{\src\uav} \bar{\rho}_{\src\uav}^2 + 1} $ and
    $\avgsnr \triangleq \avgsnr_{\gtoa} \frac{K_{\src\uav} \bar{\rho}_{\src\uav}^2 + 1}{K_{\src\uav} + 1}$.
The mixing proportion of each $m^{th}$ noncentral $\chi^2$ distribution is 
\begin{align}
\varkappa_m \triangleq \binom{M-1}{m} (\bar{\rho}_{\src\uav}^2)^m \frac{(K_{\src\uav}+1)^m}{(K_{\src\uav} \bar{\rho}_{\src\uav}^2+1)^{M-1}} (\rho_{\src\uav}^2)^{M-m-1}. \nonumber
\end{align}

Hence, based on the CDF noncentral $\chi^2$ RV's CDF in \eqref{eq:cdf_noncentralChi2}, we formulate the CDF of the G2A SNR as
\begin{align}
F_{\snr_{\gtoa}}(x) 
    =   \sum_{m=0}^{M-1} \varkappa_m 
    F_{\widetilde{\displaystyle \chi}^2}\left( 
        M - m, \lambda; \frac{x}{\avgsnr} 
    \right),~x > 0.
\label{eq:cdf_snrG2a}
\end{align}
\end{Corollary}

\vspace{-20pt}
\subsection{Statistical Characterization of RIS-assisted A2G Communication}

As outlined in \cite{AbualhayjaCL2024, YangTVT2022, AlvaradoWCL2024, LiWCL2024}, the phase-shift configuration ${\vartheta_{\ris_n} = -\angle \left[ \mathbf{h}_{\uav\ris} \right]_n - \angle \left[ \mathbf{h}_{\ris\des} \right]_n}$, for all $n \in {\cal N}$, is optimal in the absence of channel aging.
    However, adopting the aforementioned configurations in the presence of channel aging requires the current CSI $\mathbf{h}_{\uav\ris}$ and $\mathbf{h}_{\ris\des}$, which are usually unavailable in practice since ${\bf z}_{\uav\ris}$ and ${\bf z}_{\ris\des}$ are unknown.
    This inspired us to focus on PSCs that are independent of ${\bf z}_{\uav\ris}$ and ${\bf z}_{\ris\des}$.
Hereafter, we rewrite the phase-shift matrix as 
$\mathbf{\Theta}_{\ris} = \boldsymbol{\Theta}_{\ris\des}^{\sf H} \boldsymbol{\beta}_{\ris} \boldsymbol{\Theta}_{\uav\ris}$,
where
    $\boldsymbol{\beta}_{\ris} = \diag([\beta_1, \beta_2, \dots, \beta_N])$ is the RIS amplitude reflection coefficients matrix,
    ${\boldsymbol{\Theta}_{\uav\ris} = \diag( [e^{j\vartheta_{\uav\ris_1}}, \dots, e^{j\vartheta_{\uav\ris_N}}] )}$
    and ${\boldsymbol{\Theta}_{\ris\des}^{\sf H} = \diag( [e^{j\vartheta_{\ris_1\des}}, \dots, e^{j\vartheta_{\ris_N\des}}] )}$ are the phase-shift matrices of the UAV-to-RIS and RIS-to-GUE channels, respectively.
Thus, the phase-shift of the reflecting element $\forall n \in {\cal N}$ is rewritten as ${\vartheta_{\ris_n} = \vartheta_{\uav\ris_n} - \vartheta_{\ris_n\des}}$.    
Hence, we assume that the configured channel vectors 
    $\boldsymbol{\mu}_{\uav\ris} \triangleq (\boldsymbol{\beta}_{\ris})^{\frac{1}{2}} \boldsymbol{\Theta}_{\uav\ris} {\bf h}_{\uav\ris}$ and 
    $\boldsymbol{\mu}_{\ris\des} \triangleq (\boldsymbol{\beta}_{\ris})^{\frac{1}{2}} \boldsymbol{\Theta}_{\ris\des} {\bf h}_{\ris\des}$ are characterized as
\begin{align}
\boldsymbol{\mu}_{\uav\ris}
    &\mathop{=}^d   
    \mathcal{CN}\left( 
        \rho_{\uav\ris}  (\boldsymbol{\beta}_{\ris})^{\frac{1}{2}} \boldsymbol{\Theta}_{\uav\ris} \widehat{\bf h}_{\uav\ris}, 
        \bar{\rho}_{\uav\ris}^2 \boldsymbol{\beta}_{\ris} \right),
\label{eq:tilde_hUR_modelCN} \\
\boldsymbol{\mu}_{\ris\des}
    &\mathop{=}^d  
    \mathcal{CN}\left( \rho_{\ris\des} (\boldsymbol{\beta}_{\ris})^{\frac{1}{2}} \boldsymbol{\Theta}_{\ris\des}^{\sf H} \widehat{\bf h}_{\ris\des}, 
    \bar{\rho}_{\ris\des}^2 \boldsymbol{\beta}_{\ris} \right)
\label{eq:tilde_hRD_modelCN},
\end{align}
respectively. Here, \eqref{eq:tilde_hUR_modelCN} and \eqref{eq:tilde_hRD_modelCN} suggest that
\begin{itemize}
    \item Focusing solely on the phase shifts $\boldsymbol{\Theta}_{\ris\des}$ and $\boldsymbol{\Theta}_{\uav\ris}$ configurations, we can adjust the value at which the channel power is concentrated.
    \item The reflection coefficient $\boldsymbol{\beta}_{\ris}$, not only influences the value at which the channel power is concentrated, but also shapes the spread or dispersion of the channel's power distribution.
\end{itemize}

Let $\widehat{\boldsymbol{\mu}}_{\uav\ris} \triangleq (\boldsymbol{\beta}_{\ris})^{\frac{1}{2}} \boldsymbol{\Theta}_{\uav\ris} \widehat{\bf h}_{\uav\ris}$ and
    $\widehat{\boldsymbol{\mu}}_{\ris\des} \triangleq (\boldsymbol{\beta}_{\ris})^{\frac{1}{2}} \boldsymbol{\Theta}_{\ris\des}^{\sf H} \widehat{\bf h}_{\ris\des}$, 
the PDF of $\boldsymbol{\mu}_{\uav\ris}$ conditioned on $\widehat{\boldsymbol{\mu}}_{\uav\ris}$ is given by
\begin{align}
f_{\boldsymbol{\mu}_{\uav\ris} | \widehat{\boldsymbol{\mu}}_{\uav\ris}}(\mathbf{y})
    =   \frac{|\boldsymbol{\beta}_\ris|^{-1}}{\pi^N \bar{\rho}_{\uav\ris}^{2N}} 
    e^{- \frac{1}{\bar{\rho}_{\uav\ris}^2} 
        (\mathbf{y} - \rho_{\uav\ris} \widehat{\boldsymbol{\mu}}_{\uav\ris})^{\sf H} \boldsymbol{\beta}_{\ris}^{-1} (\mathbf{y} - \rho_{\uav\ris} \widehat{\boldsymbol{\mu}}_{\uav\ris})},
\label{eq:pdf_muUR}
\end{align}
for $\mathbf{y} \in \mathbb{C}^N$.

\subsubsection{Effective Air-to-Ground Channel}
    In the following, we first study the statistical characterization of the effective A2G channel through the~RIS.

\begin{Lemma}
\label{lem:lem_charFunc_Z1Z2}
Denoting $Z \triangleq \boldsymbol{\mu}_{\ris\des}^{\sf H} \boldsymbol{\mu}_{\uav\ris}$ as the effective A2G channel, the CF of $Z$ conditioned on the estimated CSI, i.e., $\hat{h}_{\uav\ris_n}$ and $\hat{h}_{\ris_n\des}$, $n = 1, 2, \dots, N$, and the phase shifts, i.e., $\vartheta_{\ris_n}$, $n = 1, 2, \dots, N$, is obtained as
%
%
\begin{align}
&\Phi_{Z|\widehat{\boldsymbol{\mu}}_{\uav\ris}, \widehat{\boldsymbol{\mu}}_{\ris\des}}(j\omega)
    =  \prod_{n=1}^{N}
    \left(
        1+ \frac{\left| \omega \right|^2}{4} \bar{\rho}_{\uav\ris}^2 \bar{\rho}_{\ris\des}^2 \beta_n^2
    \right)^{-1}
    \nonumber\\
    &\quad\times
    \exp\left\{
        - \frac{\frac{\left| \omega \right|^2}{4} \bar{\rho}_{\uav\ris}^2 \bar{\rho}_{\ris\des}^2 \beta_n^2}{1+ \frac{\left| \omega \right|^2}{4} \bar{\rho}_{\uav\ris}^2 \bar{\rho}_{\ris\des}^2 \beta_n^2}
        \left(
            \frac{\rho_{\uav\ris}^2}{\bar{\rho}_{\uav\ris}^2} 
            \left| \hat{h}_{\uav\ris_n} \right|^2  
            + \frac{\rho_{\ris\des}^2}{\bar{\rho}_{\ris\des}^2} 
            \left| \hat{h}_{\ris_n\des} \right|^2 
        \right)
    \right.
\nonumber\\
    &\qquad\qquad\qquad
    \left.
        + j \rho_{\ris\des} \rho_{\uav\ris}
        \frac{\real\left( 
            \omega^\ast \hat{h}_{\uav\ris_n} \beta_n e^{j\vartheta_{\ris_n}} \hat{h}_{\ris_n\des}
        \right)}{1+ \frac{\left| \omega \right|^2}{4} \bar{\rho}_{\uav\ris}^2 \bar{\rho}_{\ris\des}^2 \beta_n^2}
    \right\},~\omega \in \mathbb{C},
\label{eq:charFunc_Z1Z2}
\end{align}
%
\end{Lemma}
\begin{IEEEproof}
    We first derive $\Phi_{Z_1, Z_2}(j\omega_1, j\omega_2)$ in Appendix \ref{apx:proof_lem_1} as the joint CF of $Z_1 = \real(Z)$ and $Z_2 = \imag(Z)$.~Then, applying the definition of the CF of complex RVs as $\Phi_{Z}(j\omega) = \mean\{ e^{j\real(\omega^\ast Z)} \} = \Phi_{Z_1 ,Z_2}(j\omega_1, j\omega_2)$, where ${\omega = \omega_1 + j\omega_2}$, for ${\omega_1 \in \mathbb{R}}$ and ${\omega_2 \in \mathbb{R}}$, and after some mathematical manipulations, we obtain \eqref{eq:charFunc_Z1Z2}. This thus completes the proof.
\end{IEEEproof}

The CF of the effective A2G channel in Lemma \ref{lem:lem_charFunc_Z1Z2} provides the following results:
\begin{itemize}
    \item The joint CF of $Z_1$ and $Z_2$, denoted as $\Phi_{Z_1, Z_2}(j\omega_1, j\omega_2)$, is obtained as $\Phi_{Z_1, Z_2}(j\omega_1, j\omega_2) = \Phi_{Z}(j \omega_1, -\omega_2)$.
    \item The CFs of $Z_1$ and $Z_2$ are
        $\Phi_{Z_1}(j\omega_1) = \Phi_{Z_1, Z_2}(j\omega_1, 0) = \Phi_{Z}(j \omega_1)$ and 
        $\Phi_{Z_2}(j\omega_2) = \Phi_{Z_1, Z_2}(0, j\omega_2) = \Phi_{Z}(- \omega_2)$, respectively.
    \item We find that $\Phi_{Z_1}(j\omega_1) \Phi_{Z_2}(j\omega_2) \ne \Phi_{Z_1, Z_2}(j\omega_1, j\omega_2)$, which implies that $Z_1$ and $Z_2$ are correlated.
    \item The effective A2G channel $Z$ can be characterized as the sum of non-IID RVs. Specifically, we can rewrite $Z$ as $Z = \sum_{n=1}^{N} {Z^{(n)}}$, where $Z^{(1)}, \dots, Z^{(N)}$ are non-IID RVs. 
    Comparing with \eqref{eq:snr_a2g}, we determine that $Z^{(n)} = h_{\ris_n\des} \beta_n e^{j\vartheta_{\ris_n}} h_{\uav\ris_n}$.
\end{itemize}

Let $Z_1^{(n)} \triangleq \real\left( Z^{(n)} \right)$ and 
    $Z_2^{(n)} \triangleq \real\left( Z^{(n)} \right)$, we propose the following characterizations:
\begin{align}
Z_1^{(n)} &\mathop{=}^{d}
    \beta_n
    \left[ 
          \left| \frac{Z_{\uav\ris_n} + Z_{\ris_n\des}}{\sqrt{2}} + X_1^{(n)} \right|^2 
    \right.
\nonumber\\
    &\qquad\quad
    \left.
        - \left| \frac{Z_{\uav\ris_n} - Z_{\ris_n\des}}{\sqrt{2}} + Y_1^{(n)} \right|^2
    \right]
    \frac{\bar{\rho}_{\ris\des} \bar{\rho}_{\uav\ris}}{2} , \label{eq:Z1} \\
Z_2^{(n)} &\mathop{=}^{d}
    \beta_n
    \left[ 
        \left| \frac{Z_{\uav\ris_n} - j Z_{\ris_n\des}}{\sqrt{2}} + X_2^{(n)} \right|^2 
    \right.
\nonumber\\
    &\qquad\quad
    \left.
        - \left| \frac{Z_{\uav\ris_n} + j Z_{\ris_n\des}}{\sqrt{2}} + Y_2^{(n)} \right|^2
    \right]
    \frac{\bar{\rho}_{\ris\des} \bar{\rho}_{\uav\ris}}{2} , \label{eq:Z2}
\end{align}
where 
    $Z_{\uav\ris_n} = \frac{\rho_{\uav\ris}}{\bar{\rho}_{\uav\ris}} \hat{h}_{\uav\ris_n} e^{j \vartheta_{\uav\ris_n}}$,
    $Z_{\ris_n\des} = \frac{\rho_{\ris\des}}{\bar{\rho}_{\ris\des}} \hat{h}_{\ris_n\des} e^{-j \vartheta_{\ris_n\des}}$, and 
    $X_1^{(n)}$, $Y_1^{(n)}$, $X_2^{(n)}$, and $Y_2^{(n)}$ are independent and identically distributed standard complex circularly symmetric Gaussian RVs.
%
Here, we utilize the notation $\mathop{=}^d$, rather than $\mathop{\approx}^d$, due to exact matches between the right-hand sides (RHSs) and left-hand sides (LHSs) of \eqref{eq:Z1} and \eqref{eq:Z2}. 
    However, it is important to note that \eqref{eq:Z1} and \eqref{eq:Z2} do not account for correlations between $Z_1^{(n)}$ and $Z_2^{(n)}$. Nonetheless, these correlations are insignificant in characterizing the distribution of $|Z|^2$ when $N$ is sufficiently large.


%
\begin{proposition}
\label{prop:Zdis_clt}
Under channel aging, for sufficiently large $N$, the RIS-assisted A2G fading is characterized as
\begin{IEEEeqnarray}{lCl}
Z   &\mathop{\to}^d& 
    \mathcal{CN}\Big( 
        \mu_Z \!\triangleq\! {\rho}_{\ris\des} {\rho}_{\uav\ris} 
        \widehat{\bf h}_{\ris\des}^{\sf H} \boldsymbol{\Theta}_{\ris} \widehat{\bf h}_{\uav\ris}, 
        \sigma_Z^2 \!\triangleq\! 
        {\rho}_{\ris\des}^2 \bar{\rho}_{\uav\ris}^2 
        \Big\Vert \boldsymbol{\beta}_{\ris} \widehat{\mathbf{h}}_{\ris\des} \Big\Vert^2
    \nonumber\\
    &&\qquad\qquad
        + \bar{\rho}_{\ris\des}^2 {\rho}_{\uav\ris}^2 
        \Big\Vert \boldsymbol{\beta}_{\ris} \widehat{\mathbf{h}}_{\uav\ris} \Big\Vert^2
        + \bar{\rho}_{\ris\des}^2 \bar{\rho}_{\uav\ris}^2 
        \left\Vert \boldsymbol{\beta}_{\ris} \right\Vert^2_{\sf F}
    \Big). \label{eq:Z}
\end{IEEEeqnarray}
\end{proposition}

\begin{IEEEproof}
We find that
\begin{align}
Z^{(n)}_1 &\mathop{=}^d 
    \frac{\widetilde{\chi}_1^2 \left( X_1 \right) 
        -  \widetilde{\chi}_1^2 \left( X_2 \right)}{2}
    \beta_n \bar{\rho}_{\uav\ris} \bar{\rho}_{\ris\des}.
\end{align}
where ${X_1 \triangleq \frac{1}{2} \left| Z_{\uav\ris_n} + Z_{\ris_n\des} \right|^2}$ 
    and ${X_2 \triangleq \frac{1}{2} \left| Z_{\uav\ris_n} - Z_{\ris_n\des} \right|^2}$.
    
For a noncentral $\chi^2$ RV modeled as $Y \mathop{=}^d \chi_k^2(X)$, the conditional mean and conditional variance of $Y$ given $X$ are ${\mean(Y|X) = k + X}$ and ${\var(Y|X) = 2k+4X}$, respectively. 
    The total mean of $Y$ is directly computed by taking the expectation of $\mean(Y|X)$ over $X$ as $\mean(Y) = \mean(\mean(Y|X)) = k + \mean(X)$. 
Consequently, the total mean of $Z^{(n)}_1$ is computed as
\begin{align}
\mean\left[ Z^{(n)}_1 \right]
    =  \frac{\mean\left[ X_1 - X_2 \right]}{2} 
    \beta_n \bar{\rho}_{\uav\ris} \bar{\rho}_{\ris\des}.
\end{align}

The total variance of $Y$ follows the law of total variance as $\var(Y) = \mean(\var(Y|X)) + \var\left( \mean(Y|X) \right)$ \cite{GuruacharyaTWC2018}. 
Hence, the total variance of $Z^{(n)}_1$ is computed as
\begin{align}
\var\left[ Z^{(n)}_1 \right] 
    =  \frac{2 \mean\left( 1 +\! X_1 +\! X_2 \right)
        + \var\left( X_1 -\! X_2 \right)}{4} 
    \beta_n^2 \bar{\rho}_{\uav\ris}^2 \bar{\rho}_{\ris\des}^2.
\end{align}

Hence, the total mean and total variance of $Z_1$ are computed~as
\begin{align}
\mean(Z_1) &= \mean\left[ \tsum_{n=1}^{N} Z_1^{(n)} \right]
    \mathop{=}\limits^{(a)} \tsum_{n=1}^{N} \mean\left[ Z_1^{(n)} \right]
    = \mean[\real(\mu_Z)], \label{eq:mean_Z1} \\
\var(Z_1) 
    &= \var\left[ \tsum_{n=1}^{N} Z_1^{(n)} \right] 
    \mathop{=}\limits^{(b)} \tsum_{n=1}^{N} \var\left[ Z_1^{(n)} \right]
\nonumber\\
    &= \var[\real(\mu_Z)] + \mean[\sigma_Z^2], \label{eq:var_Z1}
\end{align}
respectively, where $(a)$ is based on the linearity of expectation, and $(b)$ is due to the fact that $Z^{(n)}_1$ are mutually independent, as shown in \eqref{eq:Z1}.
The mean and variance of $Z_2$ are analogous to those of $Z_1$ in \eqref{eq:mean_Z1} and \eqref{eq:var_Z1}, respectively, i.e., by simply replacing $\real(\mu_Z)$ with $\imag(\mu_Z)$.
    
    According to the CLT, $Z \mathop{\to}^d \mathcal{CN}\left( \mean[Z], \var[Z] \right)$, where $ \mean[Z] = \mean[\mu_Z]$ and $\var[Z] = \var[\mu_Z] + \mean[\sigma_Z^2]$. However, the CLT necessitates an asymptotically large $N$. 
This motivates us to explore the relationships $\sigma_Z^2 \to \mean\left[ \sigma_Z^2 \right]$ and $\mu_Z \to^d {\cal CN}\left( \mean[\mu_Z], \var[\mu_Z] \right)$, resulting in \eqref{eq:Z}. In this manner, we conclude the proof of Proposition~\ref{prop:Zdis_clt}.
\end{IEEEproof}

\subsubsection{Effective Air-to-Ground Signal-to-Noise Ratio}
\label{subsec:eff_atog_snr}

Based on Proposition \ref{prop:Zdis_clt}, we note that the distribution of $|Z|^2$ 
in the ideal scenario $\rho_{\uav\ris}^2 = \rho_{\ris\des}^2 = 1$ is simplified as
\begin{align}
    |Z|^2 \mathop{=}^d \left| \widehat{\bf h}_{\ris\des}^{\sf H} \boldsymbol{\Theta}_{\ris} \widehat{\bf h}_{\uav\ris} \right|^2,
\end{align}
which can be effectively characterized by the noncentral $\chi^2$ and generalized-$K$-squared distribution \cite{DoTCOM2021, BansalTC2023, YangTVT2022, YangTVT2020}.

When $\rho_{\uav\ris}^2 \to 0$ and $\rho_{\ris\des}^2 \to 0$, e.g., the UAV and GUE move with very high speed, the distribution of $|Z|^2$ is rewritten as $|Z|^2 \mathop{\to}^d \left( \sum_{n=1}^{N} \beta_n^2 \right) \widetilde{\chi}_1^2$. 
    In this case, the effective A2G channel exhibits Rayleigh-like fading. Hence, configuring the phase shifts $\boldsymbol{\Theta}_{\uav\ris}$ and $\boldsymbol{\Theta}_{\ris\des}$ is ineffective in highly dynamic environments. Nonetheless, increasing the number of reflecting elements can yield considerable performance gains without the need for effective PSCs.

Most importantly, when $\rho_{\uav\ris}^2, \rho_{\ris\des}^2 \in (0, 1)$, the A2G SNR can be characterized as
\begin{align}
{\snr}_{\atog} \mathop{\to}^{d} 
\avgsnr_{\atog} |Z|^2 \mathop{\to}^{d} 
    \avgsnr_{\atog} \sigma_Z^2 
    \widetilde{\chi}_1^2\left( \frac{|\mu_Z|^2}{\sigma_Z^2} \right) 
    \triangleq \avgsnr_{\atog} \sigma_Z^2 \gamma_Z,
    \label{eq:Z2_cond}
\end{align}
for sufficiently large $N$.
Here, the scale component $\sigma_Z^2$ is the linear combination of non-IID noncentral $\chi^2$ RVs, which follows the generalized noncentral $\chi^2$ distribution \cite{Mathai1992}. 
    While the exact PDF (or CDF) of the generalized noncentral $\chi^2$ distribution remains unknown, the distribution is well characterized by a noncentral $\chi^2$ distribution \cite{Mathai1992}.
    The approach followed to match the true distribution of an arbitrary RV $Y$ to the noncentral $\chi^2$ distribution is presented in Lemmas \ref{lem:cmoment_Y} and \ref{lem:momentEstm_Y} as follows. 

\begin{Lemma}
\label{lem:cmoment_Y}
For $Y \mathop{=}^d \avgsnr \widetilde{\chi}_{k}^2(X)$, the mean, the second central moment (variance), and the third central moment of $Y$ are, respectively, presented~as
\begin{subequations}
\begin{align}
\mean[Y] &= (k+\mean[X]) \avgsnr, \\
\var[Y] &= (k + \var[X] + 2\mean[X]) \avgsnr^2, \label{eq:moment_Y_b} \\
\mu_3[Y] &= \left( 2k+6\mean[X]+6\var[X] + \mu_3[X] \right) \avgsnr^3. \label{eq:mu3_Y}
\end{align}
\label{eq:moment_Y}
\end{subequations}
\vspace{-15pt}
\end{Lemma}
\begin{IEEEproof}
We adopt the LTC \cite{GuruacharyaTWC2018, brillinger1969calculation}, where the law of total variance, utilized in the proof of Proposition~\ref{prop:Zdis_clt}, and the third central moment are special cases of LTC.
    Here, the third central moment of $Y$ is derived as \cite{GuruacharyaTWC2018, brillinger1969calculation}
\begin{align}
\mu_3(Y) &= 3 \cov\left[ \mean\left( Y|X \right), \var\left( Y|X \right) \right] 
\nonumber\\
    &\qquad
    + \mean\left[ \mu_3(Y|X) \right] + \mu_3\left[ \mean\left( Y|X \right) \right],
\label{eq:mu3Y_X}
\end{align}
where the conditional moment and central moments are computed as ${\mean(Y|X) = (k+X) \avgsnr}$, ${\var(Y|X) = (1+2X) \avgsnr^2}$, and $\mu_3(Y|X) = (2+6X) \avgsnr^3$, respectively.
Furthermore, we have $\mean[\mu_3(Y|X)] = (2+6\mean[X]) \avgsnr^3$, $\mu_3[\mean(Y|X)] = \mu_3[X] \avgsnr^3$, and $\cov[\mean(Y|X), \var(Y|X)] = 2 \var[X] \avgsnr^3$.
    Plugging the foregoing results into \eqref{eq:mu3Y_X} yields \eqref{eq:mu3_Y}, completing the proof of Lemma~\ref{lem:cmoment_Y}.
\end{IEEEproof}

\begin{Corollary}
When $X$ is deterministic and ${X = \lambda}$, the mean, variance, and third central moment of $Y \mathop{=}\limits^{d} \avgsnr \widetilde{\chi}_k^2(X)$ in Lemma \ref{lem:cmoment_Y} are obtained by treating $\mean[X] = \lambda$, $\var[X] = 0$, and $\mu_3[X] = 0$, which yields
\begin{subequations}
\label{eq:moment_Yspeci}
\begin{align}
\mean[Y] &= (k+\lambda) \avgsnr, \label{eq:moment_Yspeci_a} \\
\var[Y] &= (k+ 2\lambda) \avgsnr^2, \label{eq:moment_Yspeci_b} \\
\mu_3[Y] &= \left( 2k+6\lambda \right) \avgsnr^3. \label{eq:moment_Yspeci_c}
\end{align}
\end{subequations}
\end{Corollary}

\begin{Lemma}
\label{lem:momentEstm_Y}
When matching the distribution of $Y \mathop{=}^{d} \avgsnr \widetilde{\chi}_k^2(X)$ to the noncentral $\chi^2$ distribution, denoted as $Y \mathop{\to}^{d} \avgsnr_Y \widetilde{\chi}_{k_Y}^2(\lambda_Y)$, the moment-based estimators for the parameters $k_Y$, $\lambda_Y$, and $\avgsnr_Y$ are derived as

\begin{itemize}
\item {\it Case} $\var^2[Y] > \frac{1}{2} \mean[Y] \mu_3[Y] > \frac{3}{4} \var^2[Y]$:
    \begin{align}
        \avgsnr_Y &= \frac{\var[Y]}{\mean[Y]} \frac{\var[Y] - \sqrt{\var^2[Y] - \frac{1}{2} \mean[Y] \mu_3[Y]}}{\var[Y]},
            \nonumber\\
        \lambda_Y &= \frac{\frac{\var[Y]}{\avgsnr_Y} - \mean[Y]}{\avgsnr_Y},~\textrm{and}~
        k_Y = \frac{\mean[Y]}{\avgsnr_Y} - \lambda_Y.
        \label{eq:estimators_Y}
    \end{align}
\item {\it Case} $\var^2[Y] \le \frac{1}{2} \mean[Y] \mu_3[Y]$ or $\mean[Y] \mu_3[Y] \le \frac{3}{2} \var^2[Y]$:
    \begin{align}
        \avgsnr_Y = \frac{\var[Y]}{\mean[Y]},
        k_Y = \frac{\mean^2[Y]}{\var[Y]},~\textrm{and}~\lambda_Y = 0,
        \label{eq:estimators_Y2}
    \end{align}
\end{itemize}
where $\mean[Y]$, $\var[Y]$, and $\mu_3[Y]$ are derived using \eqref{eq:moment_Y} if $X$ is an RV or using \eqref{eq:moment_Yspeci} if $X$ is deterministic.
\end{Lemma}

\begin{IEEEproof}
In order to match the distribution of $Y$ to the noncentral $\chi^2$ distribution, we solve the system of equations presented in \eqref{eq:moment_Yspeci} for $k \equiv k_Y$, $\lambda \equiv \lambda_Y$, and $\avgsnr \equiv \avgsnr_Y$. 
Applying the changes of variables $x \leftarrow \avgsnr_Y$, $y \leftarrow \avgsnr_Y k_Y$, and $z \leftarrow \avgsnr_Y \lambda_Y$ yields a new system of quadratic equations in terms of $x$, $y$, and $z$. Solving this new system yields \eqref{eq:estimators_Y} and \eqref{eq:estimators_Y2}, which concludes the proof of Lemma \ref{lem:momentEstm_Y}.
\noindent 
\end{IEEEproof}

\begin{proposition}
\label{prop:dist_sigmaZ}
The true distribution of the scale component $\sigma_Z^2$ can be characterized by the noncentral $\chi^2$ distribution as
\begin{align}
\sigma_Z^2 \mathop{=}\limits^{d} 
    \avgsnr_Z \widetilde{\chi}_{N k_{\cal Z}}^2 \left( N\lambda_{\cal Z} \right).
\end{align}
\end{proposition}

\begin{IEEEproof}
The above proposition is based on matching the true distribution of $\sigma_Z^2$ to the noncentral $\chi^2$ distribution using the method in Lemma \ref{lem:momentEstm_Y}, for $Y \equiv \sigma_Z^2$, $k_Y \equiv N k_{\cal Z}$, $\lambda_Y \equiv N \lambda_{\cal Z}$, and $\avgsnr_Y \equiv \avgsnr_Z$. It is noted that the mean, variance, and third central moment of $\sigma_Z^2$ are derived as
\begin{subequations}
\label{eq:meanVar_sigmaZ}
\begin{align}
\mean\left[ \sigma_Z^2 \right] 
    &=  \left( 1 - \rho_{\ris\des}^2 \rho_{\uav\ris}^2 \right) 
    \Vert \boldsymbol{\beta}_{\ris} \Vert_{\sf F}^2, \label{eq:meanVar_sigmaZ_a} \\
\mu_k\left[ \sigma_Z^2 \right] 
    &\mathop{=}\limits^{(a)}
    \Big[ 
        \Big( \rho_{\ris\des}^2 \bar{\rho}_{\uav\ris}^2 \Big)^k \mu_k\Big( \Big| \hat{h}_{\ris_n\des} \Big|^2 \Big)
\nonumber\\
    &\qquad
        +   \Big( \bar{\rho}_{\ris\des}^2 \rho_{\uav\ris}^2 \Big)^k \mu_k\Big( \Big| \hat{h}_{\uav\ris_n} \Big|^2 \Big)
    \Big]
    \left\Vert \boldsymbol{\beta}_{\ris}^k \right\Vert_{\sf F}^2, \label{eq:meanVar_sigmaZ_b} 
\end{align}
\end{subequations}
where $\mu_2(\cdot) \equiv \var(\cdot)$ and $(a)$ is due to the fact that the $k^{th}$ central moments of a linear combination of non-IID RVs is written as $\mu_k(a X + b Y + c) = a^k\mu_k(X) + b^k \mu_k(Y)$, for $k = 2$ and $k = 3$, where $X$ and $Y$ are non-IID and $a$, $b$, and $c$ are constants. Since $\big| \hat{h}_{\AB} \big|^2$ follows a noncentral $\chi^2$ distribution, its variance and third central moment are given by \eqref{eq:moment_Yspeci}, with $k = 1$, $\lambda = K_\AB$, and $\avgsnr= \frac{1}{K_\AB+1}$.
    This completes the proof of Proposition~\ref{prop:dist_sigmaZ}.
\end{IEEEproof}

The second case of Lemma \ref{lem:momentEstm_Y} implies that the true distribution of $\sigma^2_Z$ can be matched by the $\chi^2$ distribution with $2 N k_{\cal Z}$ d.o.f., which is identical to the Gamma distribution with shape $N k_{\cal Z}$ and unit scale. 
In general, the PDF and CDF of a $\chi^2$-distributed RV with $2 k$ d.o.f. are given by \cite{DoTCOM2021}
\begin{align}
f_{{\displaystyle \chi}^2}(k; x) &= \frac{1}{\Gamma(k)} x^{k-1} e^{-x},~x > 0, \\
F_{{\displaystyle \chi}^2}(k; x) &= \frac{1}{\Gamma(k)} \gamma(k; x),~x > 0, \label{eq:cdf_chi2}
\end{align}
respectively, where $\gamma(k; x)$ is the lower incomplete Gamma function \cite[8.35]{Gradshteyn2007} and $\Gamma(k)$ is the Gamma function \cite[8.31]{Gradshteyn2007}. 

We recall from \eqref{eq:Z2_cond} that ${|Z|^2 \mathop{\to}^d \sigma_Z^2 \gamma_Z}$ for sufficiently large $N$. 
    Here, we have ${\var[\gamma_Z] = 1 + \var[X] + 2 \mean[X] > 1}$ with ${X \equiv \frac{|\mu_Z|^2}{\sigma_Z^2}}$, which is obtained from \eqref{eq:moment_Y_b}. 
Given that $\var[\sigma_Z^2]$ is considerably less than $1$, as calculated in \eqref{eq:meanVar_sigmaZ_b}, we find that $\var[\gamma_Z]$ is significantly larger than $\var[\sigma_Z^2]$.
    Consequently, the statistical properties of $|Z|^2$ are predominantly influenced by the component $\gamma_Z = \widetilde{\chi}_1^2\left( \frac{|\mu_Z|^2}{\sigma_Z^2} \right)$ in 
    \eqref{eq:Z2_cond}. 
However, the true distribution of $|Z|^2$ is not solely determined by the individual distributions of $\sigma_Z^2$ and $\gamma_Z$ but also by the correlation between them. 
    When these components are separately matched to other distributions, the existing correlation is overlooked. 
To tackle this problem, we propose the following matching:
\begin{subequations}
\begin{align}
|Z|^2 & \mathop{\to}^d \sigma_Z^2 \snr_{\cal R},~\textrm{as}~N \gg 1, 
\end{align}
\end{subequations}
subject to $\mean\left[ |Z|^2 \right] = \mean\left[ \sigma_Z^2 \snr_{\cal R} \right]$ and $\var\left[ |Z|^2 \right] = \var\left[ \sigma_Z^2 \snr_{\cal R} \right]$, 
where $\sigma_Z^2$ and $\snr_{\cal R}$ are non-IID. 
    We highlight that $\snr_{\cal R} \mathop{\to}^d \gamma_Z$ as $\cov\left( \sigma_Z^2, \gamma_Z \right) = \mean\left( |\mu_Z|^2 \right) - \mean\left( \sigma_Z^2 \right) \mean\left( \frac{|\mu_Z|^2}{\sigma_Z^2} \right) \to 0$.

\begin{proposition}
\label{prop:dist_snrZ}
The dominant component $\snr_{\cal R}$ can be characterized by the noncentral $\chi^2$ distribution as
\begin{align}
\snr_{\cal R} 
    \mathop{=}^d 
    \avgsnr_{\cal R}
    \widetilde{\chi}_{k_{\cal R} = 1}^2\left( \lambda_{\cal R} \right).
\label{eq:snr_R}
\end{align}
\end{proposition}

\begin{IEEEproof}
The mean and variance of $|Z|^2$ are derived as $\mean\big[ |Z|^2 \big] = \mean\big[ \sigma_Z^2 \big] + \mean\big[ \big|\mu_Z\big|^2 \big]$ and 
    $\var\big[ |Z|^2 \big] = \var\big[ \sigma_Z^2 \big] + \var\big[ \big| \mu_Z \big|^2 \big] + \mean\big[ \sigma_Z^4 \big] + 2 \mean\big[ \sigma_Z^2 \big] \mean\big[ \big| \mu_Z \big|^2 \big]$, where $\mean\big[ \sigma_Z^4 \big] = \var\big[ \sigma_Z^2 \big] + \mean\big[ \sigma_Z^2 \big]$.
Since $\sigma_Z^2$ and $\snr_{\cal R}$ are statistically independent, the mean and variance of ${\snr}_{\cal R}$ are computed as
\begin{align}
\mean\big[ \snr_{\cal R} \big] 
   &=  \frac{\mean\big[ |Z|^2 \big]}{\mean\big[ \sigma_Z^2 \big]}
   = \frac{\mean\big[ \sigma_Z^2 \big] + \mean\big[ \left| \mu_Z \right|^2 \big]}{\mean\big[ \sigma_Z^2 \big]}, \label{eq:mean_snrR} \\
\var\big[ \snr_{\cal R} \big] 
   &=  \frac{\var\big[ |Z|^2 \big] - \var\big[ \sigma_Z^2 \big] \mean^2\big[ \snr_{\cal R} \big]}{\var\big[ \sigma_Z^2 \big] + \mean^2\big[ \sigma_Z^2 \big]}, \label{eq:var_snrR} 
\end{align}
where the mean and variance of $\sigma_Z^2$ are computed using \eqref{eq:meanVar_sigmaZ}. 
Then, denoting $\hat{\mu}_{\ris_n} \triangleq \hat{h}_{\ris_n\des}^\ast \hat{h}_{\uav\ris_n} e^{j\vartheta_{\ris_n}}$, the mean of $|\mu_Z|^2$ is
\begin{align}
\mean\big[ \left|\mu_Z\right|^2 \big]
    &=  \tsum_{n=1}^{N} \beta_n^2 \rho_{\ris\des}^2 \rho_{\uav\ris}^2 
        \mean\big[ \left| \hat{\mu}_{\ris_n} \right|^2 \big] 
    + 2 \beta_n \rho_{\ris\des}^2 \rho_{\uav\ris}^2
    \nonumber\\
    &\quad\times
    \tsum_{m=n+1}^N
    \beta_m
    \real\left( \mean\big[
        \hat{\mu}_{\ris_n} \hat{\mu}_{\ris_m}^\ast
    \big] \right), \label{eq:mean_muZ2} 
\end{align}
where, for $L_r(x)$ denoting the $r^{th}$ order Laguerre polynomial \cite{Gradshteyn2007}, we have \cite{BansalTC2023}
\begin{IEEEeqnarray}{lCl}
\mean\left[ \left| \hat{\mu}_{\ris_n} \right|^{2 r} \right] 
    =
    \frac{L_r(-K_{\ris\des}) L_r(-K_{\uav\ris})}{ [K_{\ris\des}+1]^r [K_{\uav\ris}+1]^r }
    r^2 \Gamma^2(r). \label{eq:prop3_widetmu_k}
\end{IEEEeqnarray}

Obtaining the exact variance of $\left|\mu_Z\right|^2$ is quite complex, which motivates us to adopt the Taylor series to derive an accurate approximation. By adopting the Taylor series, where ${\frac{\partial}{\partial \real(z)} |z|^2 = 2 \real(z)}$, and ${\frac{\partial}{\partial \real(z)^2} |z|^2 = 2}$ for a complex-valued variable $z$, we obtain 
%
\begin{align}
\var\big[ \left| \mu_Z \right|^2 \big]
    &\to 
    \var^2\left[ \mu_Z \right],
    ~\textrm{if}~\var\left[ \mu_Z \right] \gg \mean\left[ \real\left(\mu_Z\right) \right] \to 0, \nonumber \\
    &\to 4 \real\left( \mean\left[ \mu_Z \right] \right)
    \var\left[ \real\left( \mu_Z \right) \right],~\textrm{otherwise},
\end{align}
where $\var\big[ \real\big(\mu_Z\big) \big] \!\to\! \mean\big[ \big| \mu_Z \big|^2 \big] \!-\! \big| \mean\big[ \mu_Z \big] \big|^2$ if ${\imag\big( \mean\big[ \mu_Z\big] \big) \!\to\! 0}$, otherwise $\var\!\big[ \real\big(\mu_Z\big) \big] \!\to\! \frac{1}{2} \big[ \mean\big[ \big| \mu_Z \big|^2 \big] \!-\! \real\big( \mean^2\big[ \mu_Z \big] \big) \big]$. Here, 
$\mean\big[ \big| \mu_Z \big|^2 \big]$ is derived in \eqref{eq:mean_muZ2} and $\left| \mean\left[ \mu_Z \right] \right|^2$ is derived as
\begin{align}
\left| \mean\left[ \mu_Z \right] \right|^2
    &=  \left| 
        \tsum_{n=1}^{N} \beta_n \rho_{\ris\des} \rho_{\uav\ris} 
        \mean\left[ \hat{\mu}_{\ris_n} \right]
    \right|^2. \label{eq:mean_muZ} 
\end{align}

It is noted that the value of the expectations in \eqref{eq:mean_muZ2} and \eqref{eq:mean_muZ} depend on the PSC at the RIS, i.e., different PSCs yield different expectations.
    Based on the foregoing results, we obtain the mean and variance of $\snr_{\cal R}$ in \eqref{eq:mean_snrR} and \eqref{eq:var_snrR}, respectively. 
Afterwards, $\avgsnr_{\cal R}$ and $\lambda_{\cal R}$ are derived using \eqref{eq:moment_Yspeci_a} and \eqref{eq:moment_Yspeci_b} for $Y \equiv \snr_{\cal R}$, $\avgsnr \equiv \avgsnr_{\cal R}$, and $\lambda \equiv \lambda_{\cal R}$. This yields
\begin{align}
\avgsnr_{\cal R} 
    &= \mean\left[ \snr_{\cal R} \right] - \sqrt{\mean^2\left[ \snr_{\cal R} \right]-\var\left[ \snr_{\cal R} \right]}, \\
\lambda_{\cal R}
    &=  \frac{1}{\avgsnr_{\cal R}} \sqrt{\mean^2\left[ \snr_{\cal R} \right]-\var\left[ \snr_{\cal R} \right]},
\end{align}
which completes the Proof of Proposition \ref{prop:dist_snrZ}.
\end{IEEEproof}


Based on Propositions \ref{prop:dist_sigmaZ} and \ref{prop:dist_snrZ}, we characterize the RIS-assisted A2G SNR as the product of two noncentral $\chi^2$ RVs. 
    The PDF and CDF of this distribution can be represented by an infinite series of modified Bessel functions \cite{wells1962distribution}, at the cost of significantly increasing computational complexity. 
In Theorem \ref{theo:theo3}, we propose another framework to derive the distribution of the RIS-assisted A2G SNR.

\begin{Theorem}
\label{theo:theo3}
Under arbitrary PSCs, for a sufficiently large $N$, the RIS-assisted A2G SNR can be characterized by the product of two non-IID noncentral $\chi^2$ RVs. 
    In this case, the PDF of the RIS-assisted A2G SNR is obtained as
\begin{IEEEeqnarray}{lCl}
f_{\snr_\atog}(z) 
    &=&   \sum_{k_1=A_1}^{B_1} p_{{\cal K}_1}(k_1)
    \sum_{k_2=A_2}^{B_2} p_{{\cal K}_2}(k_2)
    \Big( \avgsnr_{\cal R} \avgsnr_Z \avgsnr_{\atog} \Big)^{-1}
    \nonumber\\
    &&\times 
    f_K\left( 
        Nk_{\cal Z}+k_1, k_{\cal R}+k_2; \frac{z}{\avgsnr_{\cal R} \avgsnr_Z \avgsnr_{\atog}}
    \right),
\label{eq:pdf_SNR_A2g}
\end{IEEEeqnarray}
where $p_{{\cal K}_1}(k_1) = \frac{(N \lambda_{\cal Z})^{k_1} e^{-N \lambda_{\cal Z}}}{k_1!}$ and 
$p_{{\cal K}_2}(k_2) = \frac{\lambda_{\cal R}^{k_2} e^{-\lambda_{\cal R}}}{k_2!}$ are the 
probability mass functions (PMFs) of Poisson-distributed RVs with means $N \lambda_{\cal Z}$ and $\lambda_{\cal R}$, respectively. Moreover, 
\begin{align}
f_K(a, \alpha; x) = x^{\frac{a+ \alpha}{2}-1} 
    \frac{2 K_{a-\alpha}(2\sqrt{x})}{\Gamma(a) \Gamma(\alpha)} = f_K(\alpha, a; x),
\label{eq:fK}
\end{align}
for $x > 0$ specifies the PDF of the product of two non-IID $\chi^2$-distributed RVs, which is also known as the generalized-$K$-squared distribution.
\end{Theorem}

\begin{IEEEproof}
See Appendix \ref{apx:theo3}.
\end{IEEEproof}

In Theorem \ref{theo:theo3}, the coefficients $A_i$ and $B_i$, where $i \in \{1, 2\}$, are chosen to satisfy ${\Pr( A_i \le {\cal K}_i \le B_i ) \to 1}$.
    Some special instances include $A_i = B_i = 0$ when ${\mean({\cal K}_i) = 0}$ and ${A_i = 0}$ for ${p_{{\cal K}_i}(0) > 0}$.
For large arguments, i.e., large $a$ (or $\alpha$), we propose modifying \eqref{eq:fK} as $f_K(a, \alpha; x) \to \left[ a + \frac{\alpha+(\alpha-\frac{x}{a})^2}{2} - \frac{x}{a} \right]
    \frac{f_{{\displaystyle \chi}^2}\left( \alpha; \frac{x}{a} \right)}{a}$
for $a \gg 1$. 

\begin{Corollary} \label{prop:cdf_snrA2g}
Under arbitrary PSCs, the CDF of the RIS-assisted A2G SNR is obtained by replacing $f_K(a, \alpha; x)$ in \eqref{eq:fK} with the CDF of the product of two non-IID $\chi^2$ RVs. Specifically, denoting the aforementioned CDF as $F_K(a, \alpha; x) = F_K(\alpha, a; x)$, we obtain
$F_K(a, \alpha; x)
    =   \frac{1}{\Gamma(a) \Gamma(\alpha)} 
    \textrm{G}^{2,1}_{1,3}\left(
        \left.
        \begin{smallmatrix}
            1 \\ a, \alpha, 0
        \end{smallmatrix}
        \right| x
    \right)$, $x > 0$, 
where $F_K(a, \alpha; x) \to \frac{1}{\Gamma(\alpha)} \gamma\left( \alpha; \frac{x}{a} \right) = F_{{\displaystyle \chi}^2}\left( \alpha; \frac{x}{a} \right)$ for $a \gg 1$.
\end{Corollary}

\vspace{-5pt}
\subsection{Phase-Shift Configurations}
\label{subsec:PSC}

From \eqref{eq:mean_snrR} and \eqref{eq:mean_muZ2} in Proposition \ref{prop:dist_snrZ}, we assert that the average A2G SNR, derived as
\begin{align}
\mean\left[ \snr_{\atog} \right]
    =   \avgsnr_{\atog}\left( \mean\left[ \sigma_Z^2 \right] + \mean\left[ \left| \mu_Z \right|^2 \right] \right), \label{eq:avgsnr_atog}
\end{align}
is a increasing function of $\real\left( \mean\left[ \hat{\mu}_{\ris_n} \hat{\mu}_{\ris_m}^\ast \right] \right)$, upper bounded by $\mean\left[ \left| \hat{\mu}_{\ris_n} \right| \right]
    \mean\left[ \left| \hat{\mu}_{\ris_m} \right| \right]$. Thus, configuring phase shifts to increase this value correspondingly improves the A2G SNR.
Considering PSCs where $\vartheta_{\ris_n}$ and $\vartheta_{\ris_m}$ are {\it statistically independent} for $m \ne n$, we have ${\mean\left[ \hat{\mu}_{\ris_n} \hat{\mu}_{\ris_m}^\ast \right] = \mean\left[ \hat{\mu}_{\ris_n} \right] \mean^\ast\left[ \hat{\mu}_{\ris_m} \right]}$.
    This is the case for the delayed CSI-based PSC (DPSC) where we have ${\vartheta_{\ris_n} = -\angle \hat{\mu}_{\ris_n}}$ for $n \in {\cal N}$, which does not exploit phase shift values from other reflecting elements.
It is noted that the DPSC
    yields the maximum average A2G SNR since
\begin{align}
\real\left( \mean\left[ \hat{\mu}_{\ris_n} \hat{\mu}_{\ris_m}^\ast \right] \right)
    =   \mean\left[ \left| \hat{\mu}_{\ris_n} \right| \right]
    \mean\left[ \left| \hat{\mu}_{\ris_m} \right| \right].
\end{align}

However, the DPSC comes with the trade-off of increasing computational complexity within the RIS, which becomes particularly significant as the number of RIS elements scales into the hundreds.
In practice, the RIS can utilizes other less CSI-demanding PSCs while minimizing performance degradation.
    For PSCs that disregard the information from NLoS components in the estimated CSI, we have
\begin{align}
\mean\left[ \hat{\mu}_{\ris_n} \right]
    =   \sqrt{\frac{K_{\uav\ris}}{K_{\uav\ris}+1} \frac{K_{\ris\des}}{K_{\ris\des}+1}}
    \bar{h}_{\ris_n\des}^\ast \bar{h}_{\uav\ris_n} 
    \Phi_{\vartheta_{\ris_n}}(j \omega|_{\omega = 1}),
\end{align}
where $\Phi_{\vartheta_{\ris_n}}(j \omega)$ is the CF of the phase shift $\vartheta_{\ris_n}$ for ${n \in {\cal N}}$ and $\omega \in \mathbb{R}$.
Three common PSCs that fall into this category are:
\begin{itemize}
    \item[+] Random PSCs: $\vartheta_{\ris_n}$ follows a specified distribution. The most common adopted distribution is the uniform phase-shift within the interval $[a, b] \subseteq [-\pi, \pi]$ where we obtain $\Phi_{\vartheta_{\ris_n}}(j\omega|_{\omega = 1}) = \frac{\sin b-\sin a}{b-a} + j \frac{\cos a-\cos b}{b-a}$.
    \item[+] Fixed-value (static) PSCs: $\vartheta_{\ris_n}$ is fixed and $\Phi_{\vartheta_{\ris_n}}(j) = e^{j\vartheta_{\ris_n}}$ for $n \in {\cal N}$. It is noted that $\vartheta_1 = \dots = \vartheta_{\ris_N} = 0$ depicts the communication scenario without PSCs.
    \item[+] LoS-based PSCs: $\vartheta_{\ris_n}$ is a function of the LoS component. Here, $\vartheta_{\ris_l} = - \angle \bar{h}_{\uav\ris_n} + \angle \bar{h}_{\ris_n\des}$ yields the highest expected value. 
\end{itemize}

\vspace{-10pt}
\subsection{Performance Analysis}

For performance analysis, we consider the end-to-end outage probability (eOP), defined as the probability that end-to-end SNR drops below a fixed target average SNR threshold ${\avgsnr_\textrm{th} = 2^{2 R} - 1}$, where $R$ [bps/Hz] denotes the target spectral efficiency (SE). The eOP is formulated as \cite{YangTVT2020}
\begin{align}
P_{\etoe}^\textrm{out}(\avgsnr_\textrm{th}) 
    &= \Pr\left\{ \min\left[ \snr_\gtoa, \snr_\atog \right] < \avgsnr_\textrm{th} \right\} \label{eq:eOP_sim} \\
    &\mathop{=}^{(a)}  1 - \left[ 1- F_{\snr_\gtoa}(\avgsnr_\textrm{th}) \right] 
        \left[ 1- F_{\snr_\atog}(\avgsnr_\textrm{th}) \right] \label{eq:eOP_ana},
\end{align}
where $(a)$ is the result of $\snr_\gtoa$ and $\snr_\atog$ being non-IID, with $F_{\snr_\gtoa}(x)$ given in \eqref{eq:cdf_snrG2a}, and $F_{\snr_\atog}(x)$ presented in Corollary~\ref{prop:cdf_snrA2g}.
    
Additionally, when the BS transmits with sufficiently large power, we simplify the G2A SNR CDF~as
\begin{align}
F_{\snr_\gtoa}(x)
    \to
    e^{ - \frac{M K_{\src\uav} \rho_{\src\uav}^2}{K_{\src\uav} \bar{\rho}_{\src\uav}^2+1} }
    \frac{(K_{\src\uav}+1)^M (\bar{\rho}_{\src\uav}^2)^{M-1}}{(K_{\src\uav} \bar{\rho}_{\src\uav}^2+1)^M} \frac{x}{\avgsnr_{\gtoa}},
\label{eq:pdf_snrG2A_asymp}
\end{align}
for $x > 0$, and where $(x)_n \triangleq \frac{\Gamma(x+n)}{\Gamma(x)}$ is the Pochhammer symbol. 
Moreover, when $N$ is asymptotically large, e.g., hundreds of reflecting elements, the CLT yields ${\sigma_Z^2 \!\to\! \mean\left[ \sigma_Z^2 \right]}$. Denoting 
$\Omega_{\atog} \triangleq \avgsnr_{\atog} \mean\left[ \sigma_Z^2 \right] \!=\! \frac{P_\uav}{\sigma_\des^2} (1 - \rho_{\ris\des}^2 \rho_{\uav\ris}^2) \sum_{n=1}^{N}{\ell_{\ris_n\des} \beta_n^2 \ell_{\uav\ris_n}}$,
and since the normal (Gaussian) and the noncentral $\chi^2$ distributions are related as
    $\widetilde{\chi}_{k_{\cal R}}^2 \left( \lambda_{\cal R} \right) \to^d {\cal N}\left( k_{\cal R} + \lambda_{\cal R}, k_{\cal R} + 2\lambda_{\cal R} \right)$ when $\lambda_{\cal R}$ is sufficiently large, the A2G SNR CDF can be simplified~as
\begin{IEEEeqnarray}{lCl}
F_{\snr_\atog}(x)
    \to 1 - \textrm{Q}\left( \frac{\frac{x}{\avgsnr_{\cal R} \Omega_{\atog}} - \left( k_{\cal R} + \lambda_{\cal R} \right)}{\sqrt{k_{\cal R} + 2\lambda_{\cal R}}} \right),~x>0,
\label{eq:pdf_snrA2G_asymp} 
\end{IEEEeqnarray}
where $\textrm{Q}(x)$ is the Q-function. It is noted that \eqref{eq:pdf_snrA2G_asymp} is applicable even when the maximum dimension of the RIS is non-negligible, e.g., with hundreds of reflecting elements.

\vspace{-10pt}
\section{Mobility Model}

    
%

In this section, we consider a case study where the GUE follows the Random Waypoint Mobility (RWM) model. The RWP model is adopted to capture the random and unpredictable nature of the GUE's motion \cite{ZhongTVT2023}. 
    Moreover, the mobility of the UAV is characterized by the Reference Point Group Mobility (RPGM) model to introduce a degree of deviation in the UAV's motion relative to the GUE's motion~\cite{hong1999group}.

\vspace{-10pt}
\subsection{Random Waypoint Mobility (RWM) Model}

The RWM model assumes that the GUE is initially moving from a source waypoint $\mathbf{w}^{\srcwp}_{\des}(t)$ [m] to a destination waypoint $\mathbf{w}^{\deswp}_{\des}(t)$ [m].
    Here, the waypoints are statistically IID and uniformly distributed within the same active area $\area_\des$ \cite{ZhongTVT2023}.
During the movement from $\mathbf{w}^{\srcwp}_{\des}(t)$ to $\mathbf{w}^{\deswp}_{\des}(t)$, the GUE's velocity ${\mathbf{v}_{\des}(t) \in \mathbb{R}^3}$ [m/sec] remains constant and is determined~as
\begin{align}
\mathbf{v}_{\des}(t) 
    =   v_\des(t) \frac{ \mathbf{w}^{\deswp}_{\des}(t)-\mathbf{w}_{\des}^{\srcwp}(t) }{ \left\Vert \mathbf{w}^{\deswp}_{\des}(t)-\mathbf{w}_{\des}^{\srcwp}(t) \right\Vert }
    \triangleq v_\des(t) \mathbf{r}^\rwp_{\des}(t),
    \label{eq:veloc_gue}
\end{align}
where $v_{\des}(t)$ [m/sec] is the GUE's speed, which is uniformly distributed within the interval $[v_{\min}, v_{\max}]$, where $v_{\min}$ [m/sec] and $v_{\max}$ [m/sec] are the minimum and the maximum GUE speeds, respectively, and
$\mathbf{r}_{\des}^\rwp(t) \in \mathbb{R}^{3}$ represents the GUE's Cartesian direction of movement.
    Upon arring at $\mathbf{w}^{\deswp}_\des(t)$, the GUE pauses (stays still) with a probability $p_{\textrm{pause}}$ for a duration of $\tau_{\textrm{pause}}(t)$, uniformly distributed within the interval $[\tau_{\min}, \tau_{\max}]$ [sec], where $\tau_{\min}$ and $\tau_{\max}$ are the minimum and maximum pause durations, respectively. For a probability of $1 - p_{\textrm{pause}}$, the GUE moves to a new source waypoint $\mathbf{w}^{\srcwp}_{\des}(t+\Delta\tau(t)) = \mathbf{w}^{\deswp}_{\des}(t)$ and destination waypoint $\mathbf{w}^{\deswp}_{\des}(t+\Delta\tau(t))$ following the previously presented procedure.

\vspace{-10pt}
\subsection{Reference Point Group Mobility (RPGM) Model}
In the RPGM model \cite{hong1999group}, the UAV is initially positioned uniformly within a circular area $\area_{\uav}$ centered at a predefined reference point $\mathbf{p}^{\refp}_{\uav}(t) \in \mathbb{R}^3$. 
    This reference point has the same motion as the GUE, i.e., ${\mathbf{p}_{\des}(t, \tau) = \mathbf{p}^{\refp}_{\uav}(t, \tau)}$ \cite{roy2011handbook}. 
However, due to imperfect alignment, the UAV's true motion deviates from the reference point by the motion vector $\Delta\mathbf{p}_{\uav}(t, \tau)$ following a uniform distribution. 
    Therefore, the time-varying velocity of the UAV is
$\mathbf{v}_{\uav}(t) = \lim_{\tau \to 0} \frac{\mathbf{p}_{\uav}(t, \tau)}{\tau} = \mathbf{v}_{\des}(t) + \Delta\mathbf{v}_{\uav}(t)$,
where $\Delta\mathbf{v}_{\uav}(t)$ [m/sec] is the random deviation in the UAV's velocity compared to the GUE's velocity, $\Delta v_{\des}(t) \triangleq \left\Vert \Delta\mathbf{v}_{\uav} (t) \right\Vert$ denotes the UAV's speed deviation, which is uniformly distributed within the interval $[\Delta v_{\min}, \Delta v_{\max}]$.

\vspace{-10pt}
\subsection{Adaptive Target Spectral Efficiency}
When both the GUE and the UAV are in motion, the eOP widely fluctuate over time under a fixed target SE policy.
    This fluctuation poses a challenge to maintaining a consistent and reliable communication system. 
To combat this, we introduce an adaptive target approach designed to stabilize the eOP.
    This method ensures that the eOP consistently stays below a predetermined threshold $L$, increasing robustness against the unpredictable mobility of the UAV and GUE.

The adaptive target SE, denoted as $\widehat{R}$ [bps/Hz], is obtained by solving the eOP constraint  
    $P^\textrm{out}_{\mathsf{e2e}}\left( \hat{\snr}_{\textrm{th}} \right) \le L$ for $\hat{\snr}_{\textrm{th}}$, thus $\widehat{R} = \frac{\log_2(1+\hat{\snr}_{\textrm{th}})}{2}$.
Here, we adopt the following steps \cite{nguyen2024statistical}

    i) Solve $F_{\snr_{\gtoa}}\left( \hat{\snr}_{\gtoa} \right) = L$ for $\hat{\snr}_{\gtoa}$ with $F_{\snr_{\gtoa}}(x)$ given in \eqref{eq:pdf_snrG2A_asymp}.
    
    ii) Solve $F_{\snr_{\atog}}\left( \hat{\snr}_{\atog} \right) = L$ for $\hat{\snr}_{\atog}$ with $F_{\snr_{\atog}}(x)$ given in \eqref{eq:pdf_snrA2G_asymp}.
    
    iii) Compute $\hat{\snr}_{\textrm{th}} = \min\left[ \hat{\snr}_{\gtoa}, \hat{\snr}_{\atog} \right]$, yielding 
\begin{align}
\hat{\snr}_{\textrm{th}}
    &=  \min\bigg[
        e^{ \frac{K_{\src\uav} \bar{\rho}_{\src\uav}^2+1}{M K_{\src\uav} \rho_{\src\uav}^2} }
        \left( 
            \frac{K_{\src\uav}+1}{K_{\src\uav} \bar{\rho}_{\src\uav}^2+1}
        \right)^M
        \left( \bar{\rho}_{\src\uav}^2 \right)^{M-1} \avgsnr_{\gtoa} L,
\nonumber\\
    &\qquad\qquad\quad
        \Omega_{\atog} \left( \textrm{Q}^{-1}(1 \!-\! L) \sqrt{\var\left[ \snr_{\cal R} \right]} + \mean\left[ \snr_{\cal R} \right] \right)
    \bigg],
\label{eq:adaptive_gth}
\end{align}
where $\textrm{Q}^{-1}(x)$ is the inverse Q-function, i.e., $\textrm{Q}^{-1}(\textrm{Q}(x)) = x$.

\vspace{-10pt}
\section{Results and Discussions}

\begin{figure}
    \vspace{-10pt}
    \centering
    \includegraphics[width = 0.7\linewidth]{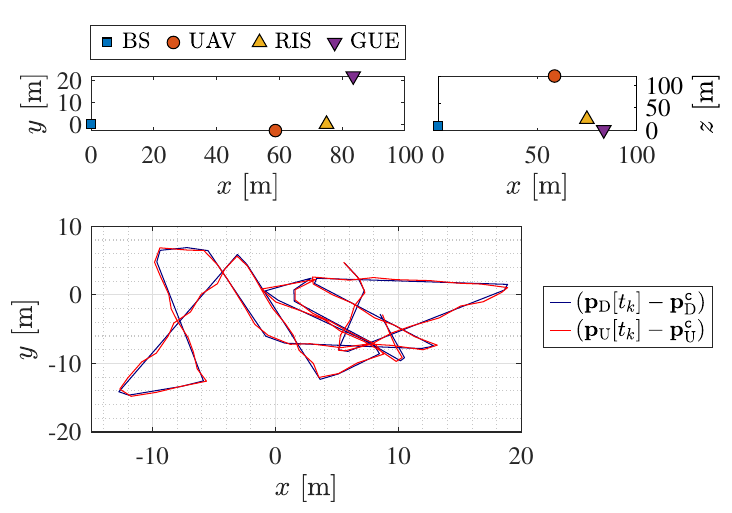}
    \caption{Illustration of the nodes' initial positions are presented in the upper figures, where $\mathbf{p}_{\src_1} = \left[ 0, 0, 10 \right]^{\sf T}$~m, 
    $\mathbf{p}_{\uav} = \left[ 56, -10, 120 \right]^{\sf T}$~m,
    $\mathbf{p}_{\ris_1} = \left[ 75, 0, 25 \right]^{\sf T}$~m, and 
    $\mathbf{p}_{\des} = \left[ 80.68, 14.15, 1.5 \right]^{\sf T}$~m. The trajectories of the GUE and the UAV are presented in the lower figure.}
    \label{fig:topology}
    \vspace{-10pt}
\end{figure}

This section presents benchmark results for the developed PDFs and CDFs of G2A and A2G SNRs, alongside the eOP and the target SE, not only to validate the correctness of the introduced Propositions and the derived Theorems, but also to provide a deeper understanding of the system behavior under channel aging.
    We highlight that the results in this section can be replicated using MATLAB. 
Moreover, for the computations of the noncentral $\chi^2$ distributions in Theorem~\ref{theo:pdf_cdf_g2asnr} and Corollary~\ref{cor:cdf_snrG2a}, we employ the built-in functions $\mathtt{ncx2cdf}$ and $\mathtt{ncx2pdf}$; while for $\chi^2$ (or Gamma) distributions, the functions $\mathtt{gamcdf}$ and $\mathtt{gampdf}$ are utilized.
    Unless stated otherwise, we present the adopted simulation parameters in Table~\ref{tab:simParameters}. 
Moreover, the BS, UAV, RIS, and GUE's positions in the global coordinate systems (i.e., MATLAB's default coordinate system) are presented in Fig.~\ref{fig:topology}.
    For practical purposes, we consider the 3GPP Urban Micro (UMi) standard to model the RIS-to-GUE channels' path loss \cite{3GPP}. Moreover, the 3GPP UMi-Airborne (UMi-AV) standard is adopted to model the BS-to-UAV (G2A) and UAV-to-RIS (A2G) path loss \cite{3gpp20173gpp}.
It is noteworthy that at a flight altitude of $120$ m adopted in Table \ref{tab:simParameters}, the LoS probability is $1.0$ \cite{3gpp20173gpp, MeerTNSM2024}.

%
%

%
\begin{table*}
    \centering
    \caption{Simulation Parameters}
    \begin{tabular}{l l l l l l}
    \toprule
        {\bf Parameter} & {\bf Value} & {\bf Parameter} & {\bf Value} & {\bf Parameter} & {\bf Value} \\
    \midrule
        Carrier frequency & $f_c = 2$ GHz \cite{3GPP2019, LiCL2023} & Offsets & $\mathbf{p}_{\uav}^\mathsf{c} = \left[ 50, 0, 120 \right]^{\sf T}$ m \cite{copley2014faa} & GUE active area & $R_\des = 20$ m \\
        Noise powers & $\sigma_\src^2 = \sigma_\uav^2 = \sigma^2$ [dBm] & & $\mathbf{p}_{\des}^\mathsf{c} = \left[ 75, 25, 1.5 \right]^{\sf T}$ m & Number of antennas & $M = 3\times 3$ antenna \\
        \multicolumn{2}{l}{\qquad\qquad\quad $\sigma^2 = N_0 + 10\log_{10} B + F$ [dBm]} & Rician factors & $K_1 = K_{\pi} = 10$ dB & Sample index & $t_k = 5000$ \\
        Noise spectral density & $N_0 = -174$ dBm/Hz \cite{LiCL2023} & GUE speed & $v_{\min} = v_{\max} \triangleq v_{\des}$ & Target SE & $R = 1$ bps/Hz \\
        Channel bandwidth & $B = 10$ MHz  \cite{3GPP, 3gpp20173gpp, 3GPP2019} & \multicolumn{2}{l}{\qquad\qquad\qquad $v_\des = 40$ m/s ($144$ km/h)} & Simulation duration & $10^5$ runs \\
        Noise figure & $F = 5$ dB & Sampling rate & $T_s = 0.01$ ms \cite{LiGLOBECOM2021} & & \\
    \bottomrule
    \end{tabular}
    \label{tab:simParameters}
    \vspace{-10pt}
\end{table*}

\begin{figure}[!h]
    \centering
    \includegraphics[width = 0.8\linewidth]{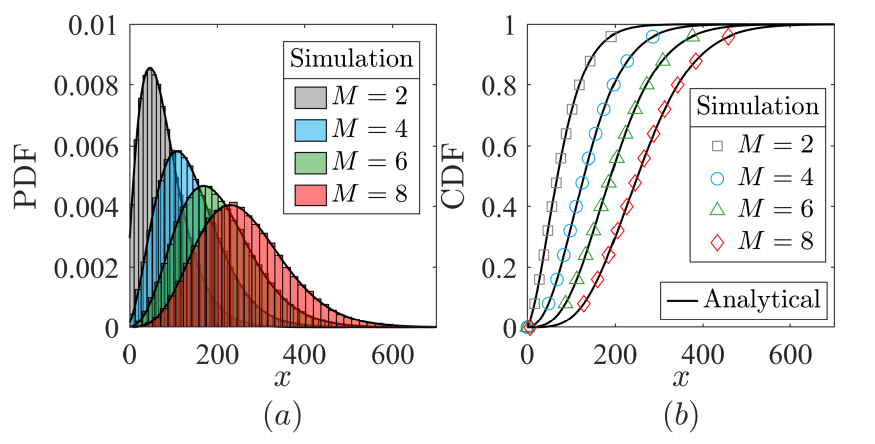}
    \caption{Simulated and analytical results of (a) the G2A SNR's PDF in \eqref{eq:pdf_snrG2a} and c) the G2A SNR's CDF in \eqref{eq:cdf_snrG2a} with different number of antennas and $P_\src = 0$ dBm.}
    \label{fig:distSnrG2a}
    \vspace{-15pt}
\end{figure}

In Fig. \ref{fig:distSnrG2a}, the simulated histograms and CDFs are obtained using the functions $\mathtt{histogram}$ and $\mathtt{ecdf}$, respectively. 
    The comparison shows a strong agreement between the analytical and simulated results for different values of $M$. 
Moreover, increasing $M$ shifts the G2A SNR's PDF and CDF to the right. Consequently, the G2A average SNR becomes higher, thereby enabling it to meet more stringent target SNR thresholds.

\begin{figure}[!h]
    \centering
    \includegraphics[width = 0.7\linewidth]{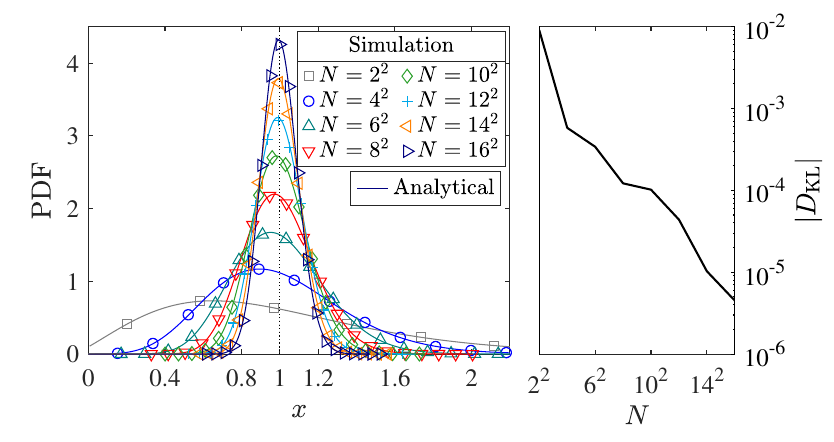}
    \caption{Simulated and analytical results of the normalized RIS-assisted A2G SNR PDF, defined as $\widetilde{\snr}_{\atog} \triangleq \frac{\snr_{\atog}}{\mean\left[ \snr_{\atog} \right]}$, with different number of reflecting elements. Here, the right figure shows the relative entropy (Kullback–Leibler divergence) between the exact A2G SNR $\snr_{\atog}$ in \eqref{eq:snr_a2g} and the proposed A2G SNR in \eqref{eq:Z2_cond}.}
    \label{fig:pdfSnrA2g}
    \vspace{-10pt}
\end{figure}

Fig. \ref{fig:pdfSnrA2g} illustrates simulated and analytical results of the A2G SNR PDF presented in Theorem \ref{theo:theo3}, where $P_{\uav} = 0$ dBm. 
    Here, we assume that the RIS utilizes the delayed-CSI based PSC and the average A2G SNR given in \eqref{eq:avgsnr_atog}.
As $N$ increases, the normalized A2G SNR PDF becomes more concentrated around the mean; specifically, the PDF gets closer to~$1$ with a decreasing variance. 
    It is noted that the Kullback–Leibler divergence, defined as $D_{\textrm{KL}} \!=\! \int_{-\infty}^\infty f_{\snr_{\atog}}(x) \log\Big( \frac{f_{\snr_{\atog}}(x)}{f'_{\snr_{\atog}}(x)} \Big) \mathrm{d} x$~\cite{DoTCOM2021}, in the right figure is produced by using MATLAB's built-in function $\mathtt{relativeEntropy}$. 
As $\left| D_{\textrm{KL}} \right| \to 0$, the A2G SNR's true PDF $f_{\snr_{\atog}}(x)$ converges to the derived PDF $f'_{\snr_{\atog}}(x)$, given in \eqref{eq:pdf_SNR_A2g}.
    Notably, the proposed characterizations for the A2G SNR based on Propositions \ref{prop:dist_sigmaZ} and \ref{prop:dist_snrZ} are accurate even when the number of reflecting elements is small, i.e., $N = 2^2$ elements, with $D_{\textrm{KL}} = 10^{-2}$,

\begin{figure}[!h]
    \vspace{-10pt}
    \centering
    \includegraphics[width = 0.7\linewidth]{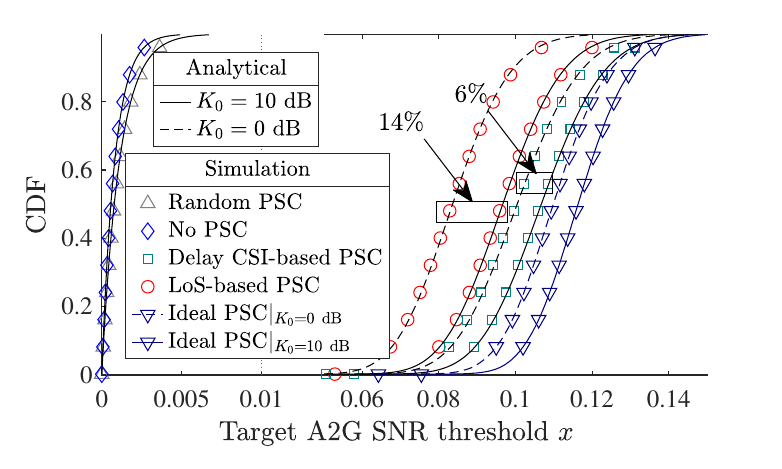}
    \caption{CDF of the A2G SNR with different PSCs, where $N = 12^2$ elements and $P_{\uav} = 0$ dBm.}
    \label{fig:cdfSnrA2g}
    \vspace{-10pt}
\end{figure}

Fig. \ref{fig:cdfSnrA2g} shows the RIS-assisted A2G SNR with the PSCs presented in Section \ref{subsec:PSC}. As can be observed, the A2G SNR's true CDF matches the analytical CDF presented in Corollary~\ref{prop:cdf_snrA2g} despite different PSCs.
    From \eqref{eq:snr_a2g_1}, we highlight that
$\snr_{\atog}
    \le  \frac{P_\uav}{\sigma_\des^2}
    \left|
        \sum_{n=1}^{N} \sqrt{{\PL}_{\ris_n\des}} 
        \left| h_{\ris_n\des} \right| \beta_n \sqrt{{\PL}_{\uav\ris_n}} 
        \left| h_{\uav\ris_n} \right|
    \right|^2$, 
where the right-hand side depicts the A2G SNR when the true CSI is known and $\vartheta_{\ris_n} = - \angle h_{\ris_n\des} - \angle h_{\uav\ris_n}$ for $n \in {\cal N}$. We refer to the aforementioned configurations for $\vartheta_{\ris_n}$ as the {\it ideal PSC} and present the corresponding A2G SNR CDF only as benchmarks.
    Furthermore, Fig. \ref{fig:cdfSnrA2g} also demonstrates that a reduction in $K_0$ from $10$ dB to $0$ dB weakens the LoS component in both UAV-to-RIS and RIS-to-GUE channels. This attenuation results in a $14\%$ performance  degradation when utilizing LoS-based PSCs as opposed to delayed CSI-based PSCs which yield a  $6\%$ loss.

\begin{figure}[!h]
    \vspace{-10pt}
    \centering
    \includegraphics[width = 0.7\linewidth]{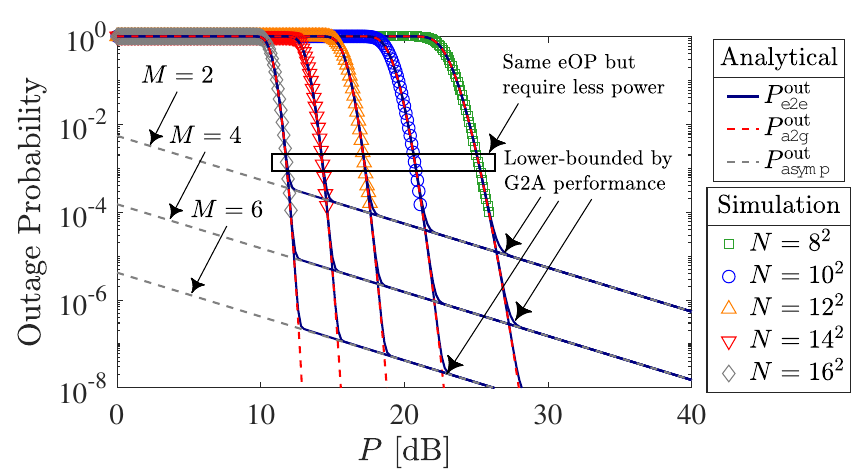}
    \caption{Simulated and analytical eOPs in \eqref{eq:eOP_sim} and \eqref{eq:eOP_ana}, respectively, with varying transmission powers, where $P^{\textrm{out}}_{\atog} = P^{\textrm{out}}_{\mathsf{e2e}}|_{ F_{\snr_\gtoa}(\avgsnr_{\textrm{th}}) = 0 }$ denotes the A2G OP.}
    \label{fig:outageProb}
    \vspace{-10pt}
\end{figure}

Next, we observe the eOP against the transmission power in Fig. \ref{fig:outageProb}, where $P_\src = P_\uav = P$ [dBm].
    It is noted that simulated results beyond $10^{-4}$ are unreliable since the number of simulation times is insufficient, e.g., $10^5$ runs in our case, hence analytical results are preferred.
The figure also shows that increasing transmission power reduces the eOP, converging to the G2A performance limit, expressed as ${P^{\textrm{out}}_{\mathsf{e2e}}(\avgsnr_{\textrm{th}}) \to F_{\snr_\gtoa}(\avgsnr_{\textrm{th}}) = \eqref{eq:pdf_snrG2A_asymp}}$. Although more transmit antennas can improve this bound, the system encounters the lower bound faster as $N$ increases.
    

\begin{figure}[!h]
    \vspace{-10pt}
    \centering
    \includegraphics[width = 0.8\linewidth]{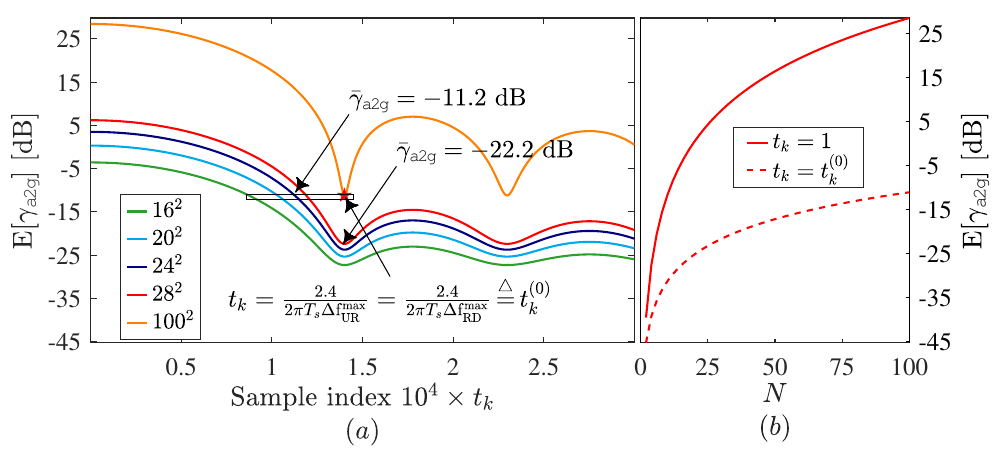}
    \caption{Average A2G SNR in decibels at $P_\uav = 0$ dBm as a function of ($a$) the sample indices $t_k$, affecting the correlation coefficients $\rho_{\uav\ris}$ and $\rho_{\ris\des}$ and ($b$) the number of RIS elements.}
    \label{fig:averageSnrA2g}
    \vspace{-10pt}
\end{figure}

Fig. \ref{fig:averageSnrA2g}(a) reveals that the average A2G SNR decreases as the sample index $t_k$ increases, due to the decreasing correlation coefficients. This decline reaches the first local minimum at $t_k = \frac{2.4}{2\pi T_s \Delta \textrm{f}^{\textrm{max}}_{\uav\ris}} = \frac{2.4}{2\pi T_s \Delta \textrm{f}^{\textrm{max}}_{\ris\des}}$, which is a decreasing function of the GUE and UAV speed, and corresponds to zero correlation, i.e., $\rho_{\uav\ris} = \rho_{\ris\des} = 0$.
    Interestingly, 
the average A2G SNR is increased to $-11.2$ dB from $-22.2$ dB when $N$ is increased from $28^2$ to $100^2$ reflecting elements. 
    This value is derived as $\mean\left[ \avgsnr_\atog \right] = \mean\left[ \avgsnr_\atog |Z|^2 \right] = \Omega_{\atog}$ based on the result in Section \ref{subsec:eff_atog_snr}.
However, a larger $N$ {\it tapers off} the increase in the average A2G SNR, as shown in Fig. \ref{fig:averageSnrA2g}(b) since the RIS's dimension also increases significantly \cite{BjornsonCM2020}.

\begin{figure}[!h]
    \vspace{-10pt}
    \centering
    \includegraphics[width = 0.7\linewidth]{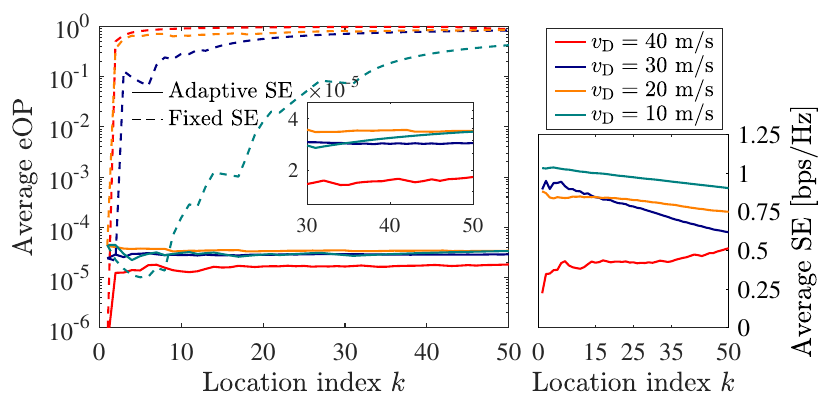}
    \caption{Average eOP and average SE over location index $k$ with and without target SE adaptation under different GUE speeds, where the target eOP in \eqref{eq:adaptive_gth} is set at $10^{-4}$ and $M = 8$ antennas.}
    \label{fig:outageProbMobility}
    \vspace{-10pt}
\end{figure}

In Fig. \ref{fig:outageProbMobility}, we assume that channel estimation is initially performed at $t_k = 0$, shown for $k \ge 1$. For subsequent locations, the average eOP at index $k$ is $\bar{P}^\textrm{out}_{\mathsf{e2e}}[k] = \frac{1}{k} P^\textrm{out}_{\mathsf{e2e}}(t_k, \avgsnr_{\textrm{th}})$, where $P^\textrm{out}_{\mathsf{e2e}}(t_k, \avgsnr_{\textrm{th}})$ denotes the eOP at the specific time instant $t_k = 5000$, for which the UAV and GUE are located at $\mathbf{p}_{\uav}[t_k]$ and $\mathbf{p}_{\des}[t_k]$, respectively.
    The figure illustrates that, with an adaptive target SE, the eOP is maintained below the desired outage level of $10^{-4}$, achieving significant stabilization, even as the GUE speed increases from $10$ m/s to $40$ m/s.
Without SE adaptation, however, the eOP is subject to considerable fluctuations, compromising communication stability and reliability over time.
    Notably, the adaptive target SE may be unnecessary for the initial ten locations, as the system already performs below the $10^{-4}$ threshold with the fixed target SE set in Table \ref{tab:simParameters}.

\vspace{-10pt}
\section{Conclusion}
In this paper, we have developed a comprehensive channel model that captures the time-varying fading, 3D mobility, Doppler shifts, and array antenna structures under the impact of channel aging. 
    Our analytical expressions for the G2A SNR PDF and CDF, based on the noncentral $\chi^2$ distribution, demonstrated that the G2A SNR follows a mixture of noncentral $\chi^2$ distributions. 
The statistical characteristics of the A2G communication is thoroughly studied, revealing that the A2G SNR can be modeled as the product of two independent noncentral $\chi^2$ RVs. 
    Furthermore, we have obtained precise PDF and CDF expressions for the product of two non-IID noncentral $\chi^2$ RVs to derive the A2G SNR distribution. 
Our results revealed that the RIS is effective against channel aging as long as the communication is carried out within a coherence time.
    Lastly, our adaptive SE method ensures stable system performance at satisfactory outage levels.

\vspace{-10pt}
\appendices
\section{Proof of Theorem \ref{theo:pdf_cdf_g2asnr}}
\label{apx:proof_g2aSNR}

Applying the Laplace transform of $\snr_\gtoa$, we obtain
${\cal L}_{\snr_{\gtoa}}(s)
    =   e^{-MK_{\src\uav}}
    \frac{(1+\Delta_1 s)^{M-1}}{(1+\Delta_2 s)^M}
    e^{MK_{\src\uav} \frac{1+\Delta_1 s}{1+\Delta_2 s} }$, 
where $\Delta_1 = \bar{\gamma}_{\src\uav}\bar{\rho}_{\src\uav}^2$ and
    $\Delta_2 = \bar{\gamma}_{\src\uav} \frac{K_{\src\uav} \bar{\rho}_{\src\uav}^2+1}{K_{\src\uav}+1}$. 
Then, the PDF of $\snr_\gtoa$ is the inverse Laplace transform of ${\cal L}_{\snr_{\gtoa}}(s)$ from $s$-domain to $x$-domain. Hence,
\begin{align}
f_{\snr_{\gtoa}}(x)
    &= e^{-MK_{\src\uav}}
    e^{-\frac{x}{\Delta_2}}
    (\Delta_1)^{M-1} (\Delta_2)^{-M}
    e^{M K_{\src\uav} \frac{\Delta_1}{\Delta_2}}
    e^{-\frac{x}{\Delta_3}}
    \nonumber\\
    &\quad\times
    \mathcal{L}^{-1}
    \Bigg\{
        \frac{s^{M-1}}{(s-\Delta_3^{-1})^M}
        e^{ MK_{\src\uav} 
        \frac{\frac{\Delta_1}{\Delta_2 \Delta_3 }}{s-\Delta_3^{-1}} };
        s, x
    \Bigg\}, 
    \label{eq_pdf_GammaSU_2} \\
    &= e^{-MK_{\src\uav}}
    e^{-\frac{x}{\Delta_2}}
    (\Delta_1)^{M-1} (\Delta_2)^{-M}
    e^{M \frac{K_{\src\uav} \Delta_1}{\Delta_2}}
    e^{-\frac{x}{\Delta_3}}
    \nonumber\\
    &\quad\times 
    \frac{{\rm d}^{M-1}}{{\rm d} x^{M-1}} 
    e^{\frac{x}{\Delta_3}}
    \mathcal{L}^{-1}
    \Bigg\{
        \frac{e^{ MK_{\src\uav} 
        \frac{\Delta_1}{\Delta_2 \Delta_3 }
        \frac{1}{s} }}{s^M};
        s, x
    \Bigg\},
    \label{eq_pdf_GammaSU_4}
\end{align}
where ${\Delta_3 \triangleq \frac{1}{\frac{1}{\Delta_1}-\frac{1}{\Delta_2}} = \frac{\Delta_1 \Delta_2}{\Delta_2-\Delta_1}}$. Here, the frequency shifting property and the derivative in the $x$-domain property of the Laplace transform are used to obtain \eqref{eq_pdf_GammaSU_2} and \eqref{eq_pdf_GammaSU_4}, respectively. 
    Using the Leibnitz's rule for higher order derivatives and \cite[Eq. (1.13.1.5)]{brychkov2008handbook}, the derivative in \eqref{eq_pdf_GammaSU_4} is derived~as
\begin{align}
\Xi(x) 
&= \frac{{\rm d}^{M-1}}{{\rm d} x^{M-1}} 
e^{\frac{x}{\Delta_3}} x^{\frac{M-1}{2}}
\frac{I_{M-1}\left( 2 \sqrt{M K_{\src\uav} 
    \frac{\Delta_1}{\Delta_2 \Delta_3} x} \right)}{ \left(M K_{\src\uav} 
    \Delta_1 \Delta_2^{-1} \Delta_3^{-1}\right)^{\frac{M-1}{2}} }
\label{eq_pdf_GammaSU_5} \\
&= \tsum_{m=0}^{M-1} \binom{M-1}{m} 
\frac{ e^{\frac{x}{\Delta_3}} x^{\frac{M-m-1}{2}} }{ \left( M K_{\src\uav} \Delta_1 \Delta_2^{-1} \Delta_3^{-1} \right)^{\frac{M-m-1}{2}} }
\nonumber\\
&\qquad\quad\times
I_{M-m-1}\left(
    2 \sqrt{M K_{\src\uav} \Delta_1 \Delta_3 \Delta_2^{-1} x}
\right).
\label{eq_pdf_GammaSU_6}
\end{align}
Substituting \eqref{eq_pdf_GammaSU_6} into \eqref{eq_pdf_GammaSU_4} and after some mathematical manipulations, we obtain \eqref{eq:pdf_snrG2a}. 

\vspace{-10pt}
\section{Proof of Lemma \ref{lem:lem_charFunc_Z1Z2}}
\label{apx:proof_lem_1} 

From the result in \cite[Eq. (3)]{MallikTC2011}, we find that $Z_1$ and $Z_2$, conditioned on $\boldsymbol{\mu}_{\uav\ris}$ and $\widehat{\boldsymbol{\mu}}_{\ris\des}$, are non-IID, and are statistically characterized as
    $Z_1 \mathop{=}^{d} 
    \mathcal{N}\left( 
        \rho_{\ris\des} 
        \real\left(
            \boldsymbol{\mu}_{\uav\ris}^{\sf H} \widehat{\boldsymbol{\mu}}_{\ris\des}
        \right),
        \frac{\bar{\rho}_{\ris\des}^2}{2} 
        \left\Vert \boldsymbol{\beta}_{\ris} \boldsymbol{\mu}_{\uav\ris} \right\Vert^2
    \right)$ and
    $Z_2 \mathop{=}^{d} 
    \mathcal{N}\left( 
        \rho_{\uav\ris}
        \imag\left(
            \boldsymbol{\mu}_{\uav\ris}^{\sf H} \widehat{\boldsymbol{\mu}}_{\ris\des}
        \right),
        \frac{\bar{\rho}_{\ris\des}^2}{2} 
        \left\Vert \boldsymbol{\beta}_{\ris} \boldsymbol{\mu}_{\uav\ris} \right\Vert^2
    \right)$,
%
%
respectively. Hence, the joint CF of $Z_1$ and $Z_2$, conditioned on $\boldsymbol{\mu}_{\uav\ris}$ and $\widehat{\boldsymbol{\mu}}_{\ris\des}$, is therefore expressed as
\begin{align}
&\Phi_{Z_1,Z_2|\boldsymbol{\mu}_{\uav\ris}, \widehat{\boldsymbol{\mu}}_{\ris\des}}(j\omega_1, j\omega_2)
    =   \exp\left[
        \rho_{\ris\des} 
        \left(
            j\omega_1 \real\left( \boldsymbol{\mu}_{\uav\ris}^{\sf H} \widehat{\boldsymbol{\mu}}_{\ris\des} \right)
        \right.
    \right.
\nonumber\\
&\quad
\left.
    \left.
        + j\omega_2 \imag\left( \boldsymbol{\mu}_{\uav\ris}^{\sf H} \widehat{\boldsymbol{\mu}}_{\ris\des} \right)
    \right)
        - \frac{\omega_1^2 + \omega_2^2}{4} \bar{\rho}_{\ris\des}^2 
        \left\Vert \boldsymbol{\beta}_{\ris} \boldsymbol{\mu}_{\uav\ris} \right\Vert^2
\right].\!\! \label{eq:cdf_Z1Z2}
\end{align}

Taking the expectation of $\eqref{eq:cdf_Z1Z2}$ over $\boldsymbol{\mu}_{\uav\ris}$, conditioned on $\widehat{\boldsymbol{\mu}}_{\uav\ris}$, based on the PDF of $\boldsymbol{\mu}_{\uav\ris}$ in \eqref{eq:pdf_muUR} yields
\begin{align}
&\Phi_{Z_1,Z_2|\widehat{\boldsymbol{\mu}}_{\uav\ris}, \widehat{\boldsymbol{\mu}}_{\ris\des}}(j\omega_1, j\omega_2)
=   \frac{|\boldsymbol{\beta}_\ris|^{-1}}{\pi^N \bar{\rho}_{\uav\ris}^{2N}}  
    \exp\left[ - \frac{{\rho}_{\uav\ris}^2}{\bar{\rho}_{\uav\ris}^2} 
        \widehat{\boldsymbol{\mu}}_{\uav\ris}^{\sf H} \boldsymbol{\beta}_\ris^{-1} \widehat{\boldsymbol{\mu}}_{\uav\ris} \right]
\nonumber\\
&\quad\times
    \int_{\mathbf{y} \in \mathbb{C}^N}
    \exp\bigg[
        - \mathbf{y}^{\sf H} \boldsymbol{\Sigma} \mathbf{y}
        +   \frac{\rho_{\uav\ris} }{\bar{\rho}_{\uav\ris}^2}
            2 \real\left( \mathbf{y}^{\sf H} \boldsymbol{\beta}_{\ris}^{-1} \widehat{\boldsymbol{\mu}}_{\uav\ris} \right)
\nonumber\\
&\qquad\qquad\qquad\qquad
        + j2 \rho_{\ris\des} \real\left( 
            \frac{\omega_1 - j\omega_2}{2} \mathbf{y}^{\sf H} \widehat{\boldsymbol{\mu}}_{\ris\des}
        \right)
    \bigg] {\rm d} \mathbf{y},
\end{align}
where $\boldsymbol{\Sigma} \triangleq \frac{\omega_1^2+\omega_2^2}{2}\bar{\rho}_{\ris\des}^2 \boldsymbol{\beta}_{\ris} + \frac{1}{\bar{\rho}_{\uav\ris}^2} \boldsymbol{\beta}_{\ris}^{-1}$ is invertible and symmetric. Then, based on \cite[Eq. (8)]{MallikTC2011}, we obtain the following solution for the complex Gaussian integrals
\begin{align}
&\frac{|\boldsymbol{\Sigma}|}{\pi^N}
\int_{\mathbf{x} \in \mathbb{C}^{N}}
    \exp\left[
        \mathbf{x}^{\sf H} \boldsymbol{\Sigma} \mathbf{x} 
        + 2 \real\left( \mathbf{x}^{\sf H} \mathbf{b} \right)
        + j 2 \real\left( \mathbf{x}^{\sf H} \mathbf{c} \right)
    \right] {\rm d} \mathbf{x} 
\nonumber\\
&\qquad
    =   \exp\left[ 
        \left\Vert \boldsymbol{\Sigma}^{-\frac{1}{2}} \mathbf{b} \right\Vert^2
        - \left\Vert \boldsymbol{\Sigma}^{-\frac{1}{2}} \mathbf{c} \right\Vert^2
        + j \real(\mathbf{b}^{\sf H} \boldsymbol{\Sigma}^{-1} \mathbf{c})
    \right],
\end{align}
and since $\omega = \omega_1 + j\omega_2$, we have 
\begin{align}
&\Phi_{Z|\widehat{\boldsymbol{\mu}}_{\uav\ris}, \widehat{\boldsymbol{\mu}}_{\ris\des}}(j\omega)
    =   \frac{\left| \boldsymbol{\Sigma} \boldsymbol{\beta}_{\ris} \right|^{-1}}{\bar{\rho}_{\uav\ris}^{2N}}
    \exp\bigg[\!
    -   \frac{ \rho_{\uav\ris}^2}{ \bar{\rho}_{\uav\ris}^2 } 
        \left\Vert \widehat{\bf h}_{\uav\ris} \right\Vert^2 
\nonumber\\
    &\quad
    +   \frac{\rho_{\uav\ris}^2}{\bar{\rho}_{\uav\ris}^4}
        \Big\Vert \boldsymbol{\Sigma}^{-\frac{1}{2}} \boldsymbol{\beta}_{\ris}^{\frac{1}{2}} \widehat{\bf h}_{\uav\ris} \Big\Vert^2
        -   \frac{|\omega|^2}{4}
        \rho_{\ris\des}^2
        \Big\Vert \boldsymbol{\Sigma}^{-\frac{1}{2}} \boldsymbol{\beta}_{\ris}^{\frac{1}{2}} \widehat{\bf h}_{\ris\des} \Big\Vert^2
\nonumber\\
    &\quad
    + j \frac{\rho_{\ris\des} \rho_{\uav\ris}}{\bar{\rho}_{\uav\ris}^2}  
        \real\Big(
            \widehat{\boldsymbol{\mu}}_{\uav\ris}^{\sf H} 
            \boldsymbol{\Sigma}^{-1} \boldsymbol{\beta}_{\ris}^{-1} \widehat{\boldsymbol{\mu}}_{\ris\des}
            \omega
        \Big)
    \bigg],~\omega \in \mathbb{C}.
\end{align}

After some mathematical manipulations, we obtain \eqref{eq:charFunc_Z1Z2}. This completes the proof of Lemma \ref{lem:lem_charFunc_Z1Z2}.

\vspace{-10pt}
\section{Proof of Theorem \ref{theo:theo3}}
\label{apx:theo3} 
The PDF of $|Z|^2$ is derived as $f_{|Z|^2}(z) = \mean\left[ \frac{1}{\sigma_Z^2} f_{\snr_{\cal R}}\left( \frac{z}{\sigma_Z^2} \right) \right]$. 
    By adopting the identity $\mean[g(X)] = \int_x g(x) f_X(x) \mathrm{d} x$ for the RV $X$ with PDF $f_X(x)$, we obtain
\begin{align}
f_{|Z|^2}(z)  
    =  \int_{0}^\infty \frac{ f_{\snr_{\cal R}}\left( {z}\big/{x} \right)}{x} f_{\sigma_Z^2}(x) \mathrm{d} x. \label{eq:apx_cdf_Z2_1}
\end{align}

We note that a noncentral $\chi^2$ RV can be characterized as a central $\chi^2$ RV with Poisson-distributed d.o.f. as $\widetilde{\chi}_k^2(\lambda) \mathop{=}^d \widetilde{\chi}_{{\cal K}+k}^2$, where ${\cal K}$ follows a Poisson distribution with mean $\lambda$. 
    Applying this property, we obtain $\sigma_Z^2 \mathop{=}^d \avgsnr_{Z} \widetilde{\chi}_{{\cal K}_1 + N k_{\cal Z}}^2$, where ${\cal K}_1$ follows a Poisson distribution with mean $N \lambda_{\cal Z}$. 
Hence, the PDF of $\sigma_Z^2$ is rewritten as 
    $f_{\sigma_Z^2}(x)
    =   \sum_{k_1 = A_1}^{B_1}
        \frac{p_{{\cal K}_1}(k_1)}{\avgsnr_{Z}} 
    f_{{\displaystyle \chi}^2}\left( N k_{\cal Z} + k_1; \frac{x}{\avgsnr_{Z}} \right)$ for $x > 0$. Similarly, we obtain $\snr_{\cal R} \mathop{=}^d \avgsnr_{\cal R} \widetilde{\chi}_{{\cal K}_2 + k_{\cal R}}^2$, where ${\cal K}_2$ follows Poisson distribution with mean $\lambda_{\cal R}$.
Moreover, the PDF of $\snr_{\cal R}$ can be rewritten as $f_{\snr_{\cal R}}(x)
    =   \sum_{k_2 = A_2}^{B_2}  
        \frac{p_{{\cal K}_2}(k_2)}{\avgsnr_{\cal R}}
    f_{{\displaystyle \chi}^2}\left( k_{\cal R} + k_2; \frac{x}{\avgsnr_{\cal R}} \right)$ for $x > 0$. 

Plugging the foregoing results into \eqref{eq:apx_cdf_Z2_1}, then applying \cite[Eq. (3.471.9)]{Gradshteyn2007} and the scaling property $f_{\snr_\atog}(z) = \avgsnr_{\atog}^{-1} f_{|Z|^2}\left( \avgsnr_{\atog}^{-1} z \right)$, we obtain \eqref{eq:pdf_SNR_A2g}, which concludes the proof of Theorem \ref{theo:theo3}.

\vspace{-10pt}
\bibliographystyle{IEEEtran}
\bibliography{References}

\end{document}